\documentclass[]{aa}
\usepackage{graphicx,natbib}

\def\jh{\mbox{$\rm (J-H)$}}

\def\jk{\mbox{$\rm (J-K_s)$}}
\def\mMJ{\mbox{$\rm (m-M)_J$}}
\def\mMo{\mbox{$\rm (m-M)_O$}}
\def\ebv{\mbox{$\rm E(B-V)$}}
\def\ejh{\mbox{$\rm E(J-H)$}}
\def\rc{\mbox{$\rm R_{C}$}}
\def\rl{\mbox{$\rm R_{RDP}$}}

\def\ms{\mbox{$\rm M_\odot$}}
\def\ds{\mbox{$\rm d_\odot$}}

\def\Rgc{\mbox{$\rm R_\odot$}}
\def\dgc{\mbox{$\rm R_{GC}$}}
\def\xgc{\mbox{$\rm x_{GC}$}}
\def\ygc{\mbox{$\rm y_{GC}$}}
\def\zgc{\mbox{$\rm z_{GC}$}}
\def\rx{\mbox{$\rm R_{ext}$}}

\def\jj{\mbox{$\rm J$}}
\def\hh{\mbox{$\rm H$}}
\def\ks{\mbox{$\rm K_s$}}
\def\aV{\mbox{$\rm A_V$}}
\def\ns{\mbox{$\rm N_{1\sigma}$}}
\def\no{\mbox{$\rm N_{obs}$}}
\def\nc{\mbox{$\rm N_{cl}$}}
\def\sFS{\mbox{$\rm\sigma_{FS}$}}
\def\fsU{\mbox{$\rm FS_{unif}$}}

\begin{document}

\title{Investigating the borderline between a young star cluster and a small stellar association:
a test case with Bochum\,1}

\author{E. Bica\inst{1} \and C. Bonatto\inst{1} \and C.M. Dutra\inst{2}}

\offprints{C. Bonatto}

\institute{Universidade Federal do Rio Grande do Sul, Departamento de Astronomia\\
CP\,15051, RS, Porto Alegre 91501-970, Brazil\\
\email{charles@if.ufrgs.br, bica@if.ufrgs.br}
\mail{charles@if.ufrgs.br}\\
\and
Universidade Federal do Pampa - UNIPAMPA, Centro de Ci\^encias da Sa\'ude\\
Rua Domingos de Almeida, 3525,
Bairro S\~ao Miguel, Uruguaiana 97500-009, RS, Brazil\\
\email{cmdutra@gmail.com} }

\date{Received --; accepted --}

\abstract
{Usually, a loose stellar distribution can be classified as an OB stellar
group, an association, or a young open cluster. We make use of comparisons with 
the typical OB association Vul\,OB1}
{In the present paper we discuss the nature of Bochum\,1, a typical example of an object
affected by the above classification problem.}
{Field-decontaminated 2MASS photometry is used to analyse Colour-Magnitude Diagrams (CMDs) and
stellar radial density profiles (RDPs) of the structures present in the region of Bochum\,1.}
{The field-decontaminated CMD of Bochum\,1 presents main sequence (MS) and
pre-main sequence (PMS) stars. We report two new small angular-size, compact young clusters and 
one embedded cluster in the area of Bochum\,1. Vul\,OB1 harbours the young open cluster NGC\,6823 
and the very compact embedded cluster Cr\,404. The Vul\,OB1 association includes the H\,II region
Sh2-86, and its stellar content is younger ($\approx3$\,Myr) than that of Bochum\,1
($\approx9$\,Myr), which shows no gas emission. Bochum\,1 harbours one of the newly
found compact clusters as its core. The RDP of Bochum\,1 is irregular and cannot be fitted by
a King-like profile, which suggests important erosion or dispersion of stars from a
primordial cluster. Similarly to Bochum\,1, the decontaminated CMD of NGC\,6823 presents conspicuous
MS and PMS sequences. Taken separately, RDPs of MS and PMS stars follow a King-like profile. The
core shows an important excess density of MS stars that mimics the profile of a post-core collapse 
cluster. At such young age, it can be explained by an excess of stars formed in the prominent core.}
{The present study suggests that Bochum\,1 is a star cluster fossil remain that might be dynamically
evolving into an OB association. Bochum\,1 can be a missing link connecting early star
cluster dissolution with the formation of low-mass OB associations.}

\keywords{({\it Galaxy}:) open clusters and associations; {\it Galaxy}: structure}

\titlerunning{Bochum\,1: stellar association or OC?}

\maketitle

\section{Introduction}
\label{intro}

Associations are loose stellar systems that may contain as much as 2600 stars, as in Cyg\,OB2 
(\citealt{Albacete02}; \citealt{Kn00}). Although sharing a common origin and moving approximately 
in the same direction through the Galaxy, the member stars are gravitationally unbound. This 
definition encompasses a wide variety of objects, from the extended OB associations in spiral 
arms (\citealt{Blaauw64}) to the post-T Tauri associations in the Solar neighbourhood (e.g. 
\citealt{Torres00}). OB associations can be observed over a wide range of distances from the Sun, 
from the relatively nearby ($\approx140$\,pc) Scorpius-Centaurus Association (\citealt{MA01}) to 
the sparse and large ($\sim400$\,pc) associations in the Large Magellanic Cloud and Andromeda 
(\citealt{EfEl98}).

Most field stars appear to have been formed in stellar groups of different kinds,
the OB associations in particular (Gomes et al. 1993; \citealt{Massey95}). The rapid early gas removal
is an efficient mechanism to drive cluster stars into the field and dissolve most of the
very young star clusters in a time scale of $10-40$\,Myr, depending on cluster mass and star-formation
efficiency (e.g. \citealt{GoBa06}). As a consequence of this infanticide, only about 5\%
(\citealt{LL2003}) of the embedded clusters are able to dynamically evolve into bound open clusters
(OCs).

The above aspects raise the fundamental issue of the detection of dispersed cluster debris, and the
distinction (if possible) between genuine small associations from dispersing debris. The present
scenario, where early-cluster disruption is favoured, may provide clues to the dispersion of large 
star-forming complexes into the field, and remaining young OC families (\citealt{Piskunov06}; 
\citealt{laF08}).

Bochum\,1 was defined by \citet{Moffat75} as a group of 8 OB stars from the catalogue of Southern
Luminous Stars (LS) by \citealt{StepSand71}. \citet{Moffat75} carried out photoelectric observations
with the ESO La Silla Bochum telescope, hence the object designation. They pointed out that the stars
stand out from the background, and the Colour-Magnitude Diagram (CMD) indicated a common reddening and
distance. They derived $\ebv=0.55\pm0.06$, $\ds=4.06$\,kpc and an O7 turn-off. Bochum\,1 is located at
$\ell=193.43^\circ$, $b=+3.40^\circ$, and $\rm\alpha(J2000)=6^h25^m30^s$ and
$\rm\delta(J2000)=19^\circ46\arcmin00\arcsec$.

\citet{Yadav03} observed Bochum\,1 with CCD photometry and referred to it as an  OB association. They
found that part of the bright stars in the area have a common proper motion, and a mass function that,
within uncertainties, is comparable to that of \citet{Salpeter55}. They derived $\ebv=0.47\pm0.10$,
$\ds=2.8$\,kpc, and an age of 10\,Myr. Bochum\,1 is included as a star cluster in the WEBDA\footnote{\em
http://www.univie.ac.at/webda - \citet{Merm03}} database, which shows the values of \citet{Yadav03}.

Recently, \citet{FSRcat} presented a catalogue of star cluster candidates detected as stellar
overdensities in the 2MASS\footnote{The Two Micron All Sky Survey --- {\em
www.ipac.caltech.edu/2mass/releases/allsky/ }} catalogue. FSR\,911 was considered to be the same
object as Bochum\,1, but the positions are not coincident (Fig.~\ref{fig1}).

In this context, the following interesting questions arise: Is Bochum\,1 a missing link between star
clusters and stellar associations? A dispersing fossil remain of a star cluster where scattered 
early-type stars and a remnant core are observed? Or a proto association that will eventually show up
only scattered stars?

In a recent series of studies we coupled the classical CMD method with the analysis of the stellar
radial density profile (RDP) to obtain intrinsic cluster astrophysical parameters and, in uncertain
cases, to establish the nature of the objects (e.g. \citealt{BB08}; \citealt{ProbFSR}; \citealt{OldOCs};
\citealt{BB07}; \citealt{N4755}). Field decontamination was crucial in all these studies.

A previous discussion on the connection between massive young star cluster and a massive OB association
was carried out by \citet{Kn00} with Cyg\,OB2. In the present study we apply these methods to Bochum\,1 
and other low-mass stellar systems in the area. We also analyse NGC\,6823 in Vul\,OB1, for comparison 
purposes among young stellar systems.

\begin{table*}
\caption[]{General data on the star cluster/associations}
\label{tab1}
\renewcommand{\tabcolsep}{1.05mm}
\renewcommand{\arraystretch}{1.25}
\begin{tabular}{ccccccccc}
\hline\hline
$\ell$&$b$&$\alpha(2000)$&$\delta(2000)$&\multicolumn{2}{c}{Diameter}&\rx&Designation&Comments\\
\cline{5-6}
($^\circ$)&($^\circ$)&(hms)&($^\circ\,\arcmin\,\arcsec$)&(\arcmin)&(pc)&(\arcmin)\\
(1)&(2)&(3)&(4)&(5)&(6)&(7)&(8)&(9)\\
\hline
\multicolumn{9}{c}{The region of Bochum\,1}\\
\hline
192.44 & $+$3.41 & 6:25:30 & $+$19:46:00 & 26.0$^a$ & 32$^d$ &60&Bochum\,1\\
192.30 & $+$3.36 & 6:25:00 & $+$19:52:03 & 14.0$^b$ & 17$^e$ &30& FSR\,911\\
192.64 & $+$3.98 & 6:25:03 & $+$19:53:00 &  2.5$^c$ & 3.1$^e$ &30& Faint-star clump & Core of FSR\,911?\\
192.31 & $+$3.36 & 6:25:01 & $+$19:50:55 &  1.0$^c$ & 1.3$^d$ &30 &New\,Cluster\,1 & Core of FSR\,911? Includes LS46 and LS47\\
192.43 & $+$3.40 & 6:25:28 & $+$19:45:10 &  0.6$^c$ & 1.2$^e$ &&New\,Cluster\,2 & Core of Bochum\,1? Includes LS51\\
192.17 & $+$3.41 & 6:24:55 & $+$19:59:59 &  1.0$^c$ & 1.3$^d$ &30&New\,Cluster\,3 & Core of FSR\,911? Includes IRAS\,06219+2001\\
\hline
\multicolumn{9}{c}{The region of NGC\,6823}\\
\hline
59.40 & $-$0.14 & 19:43:09 & $+$23:17:58 & 10.0 & 5.8$^d$ &30&NGC\,6823,Cr\,405,OCl-124\\
59.14 & $-$0.11 & 19:42:28 & $+$23:05:13 &  0.8 & 0.5$^d$ &20&Cr\,404,OCl-122 & Embedded cluster, includes IRAS\,19403+2258\\
\hline
\end{tabular}
\begin{list}{Table Notes.}
\item Cols.~1-4: Central coordinates. Angular diameters (Col.~5) from: (a) - the catalogue of
\citet{Dias02}; (b) - twice the tidal radius of \citet{FSRcat}; (c) - estimated in the present 
study from 2MASS \ks\ images. Absolute apparent diameters (Col.~6) computed
with distances from the Sun from (d) - this work (Table~\ref{tab4}); (e) - assuming the same
distance as that of Bochum\,1. Col.~7: 2MASS extraction radius. 
\end{list}
\end{table*}

This paper is structured as follows. In Sect.~\ref{targets} we provide fundamental data on Bochum\,1
and the other relevant stellar clusterings in its field, build the 2MASS CMDs and RDPs, and derive
fundamental and structural parameters. In Sect.~\ref{N6823} we apply the above methods to the template
young OC NGC\,6823. In Sect.~\ref{MF} we build mass functions and compute the stellar
content. In Sect.~\ref{Disc} we discuss whether Bochum\,1 is a genuine association or a dispersed young
star cluster (or more than one). Concluding remarks are given in Sect.~\ref{Conclu}.

\section{Stellar clusterings in the area of Bochum\,1}
\label{targets}

\begin{figure}
\resizebox{\hsize}{!}{\includegraphics{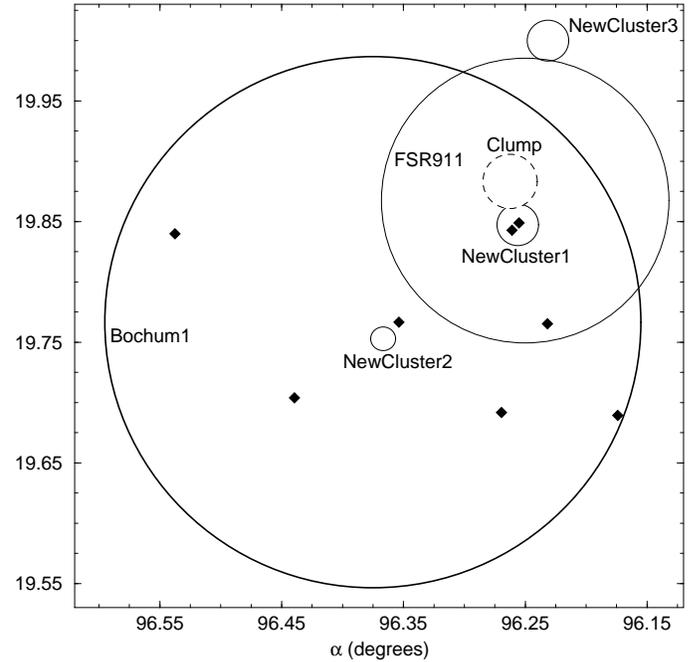}}
\caption{Schematic view of a $30\arcmin\times30\arcmin$ field centred on Bochum\,1. Objects in
this field are centred in the coordinates given in Table~\ref{tab1}, and are shown with the
angular diameters given in col.~5. The stars with spectral type determined by \citet{Moffat75}
are shown as diamonds.}
\label{fig1}
\end{figure}

Literature positions and the large angular dimensions of Bochum\,1 and FSR\,911 (Table~\ref{tab1})
are shown schematically in Fig.~\ref{fig1}. At first sight, the angular separation and dimensions 
suggest two large angular-size objects in the area. Inspection of wide-field 
XDSS\footnote{Extracted from the Canadian Astronomy Data Centre (CADC), at \em http://cadcwww.dao.nrc.ca/} 
images of Bochum\,1 and FSR\,911, shows no evident stellar concentration like an OC, nor gas emission.

%

\begin{figure*}
\begin{minipage}[b]{0.50\linewidth}
\includegraphics[width=\textwidth]{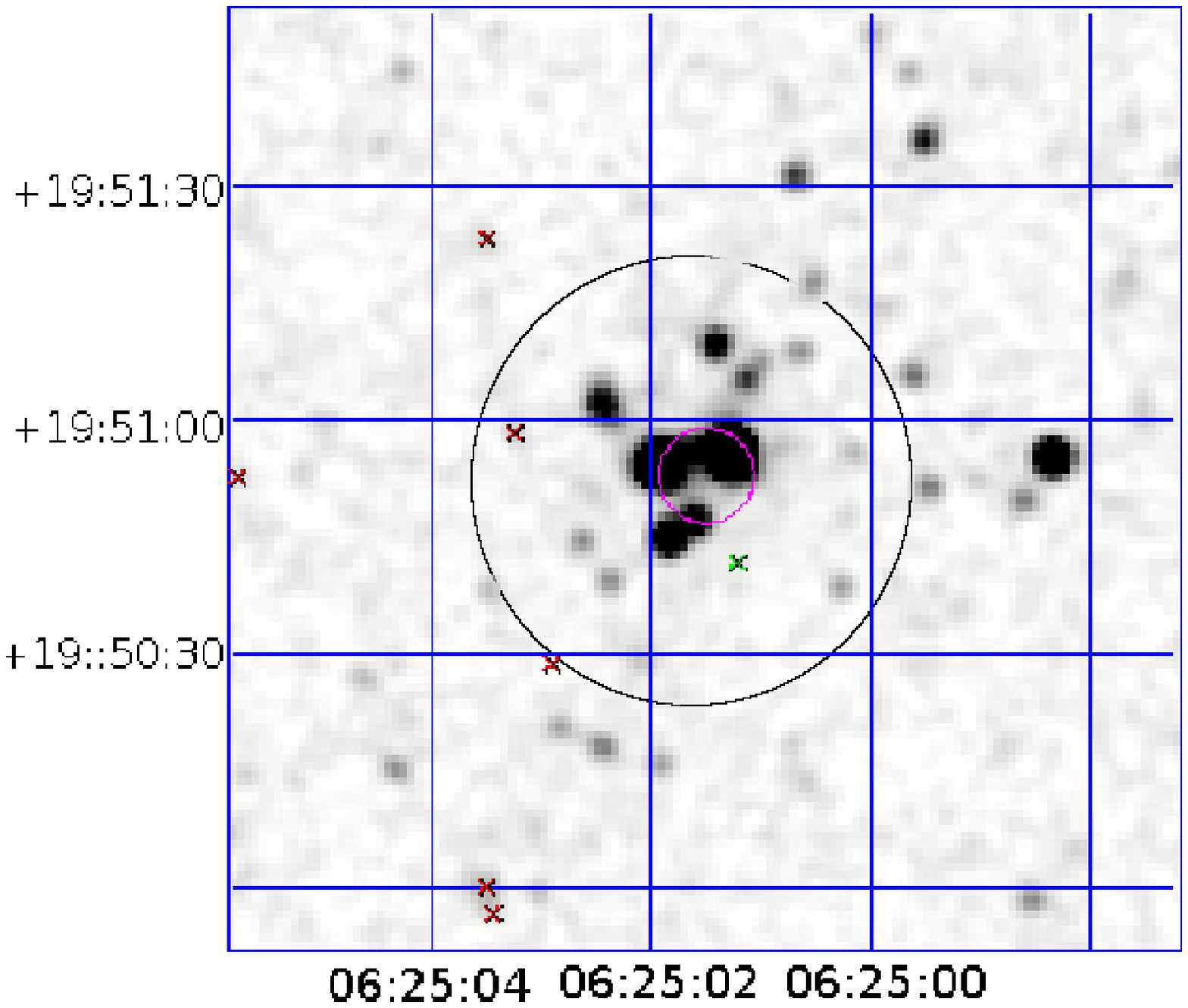}
\end{minipage}\hfill
\begin{minipage}[b]{0.50\linewidth}
\includegraphics[width=\textwidth]{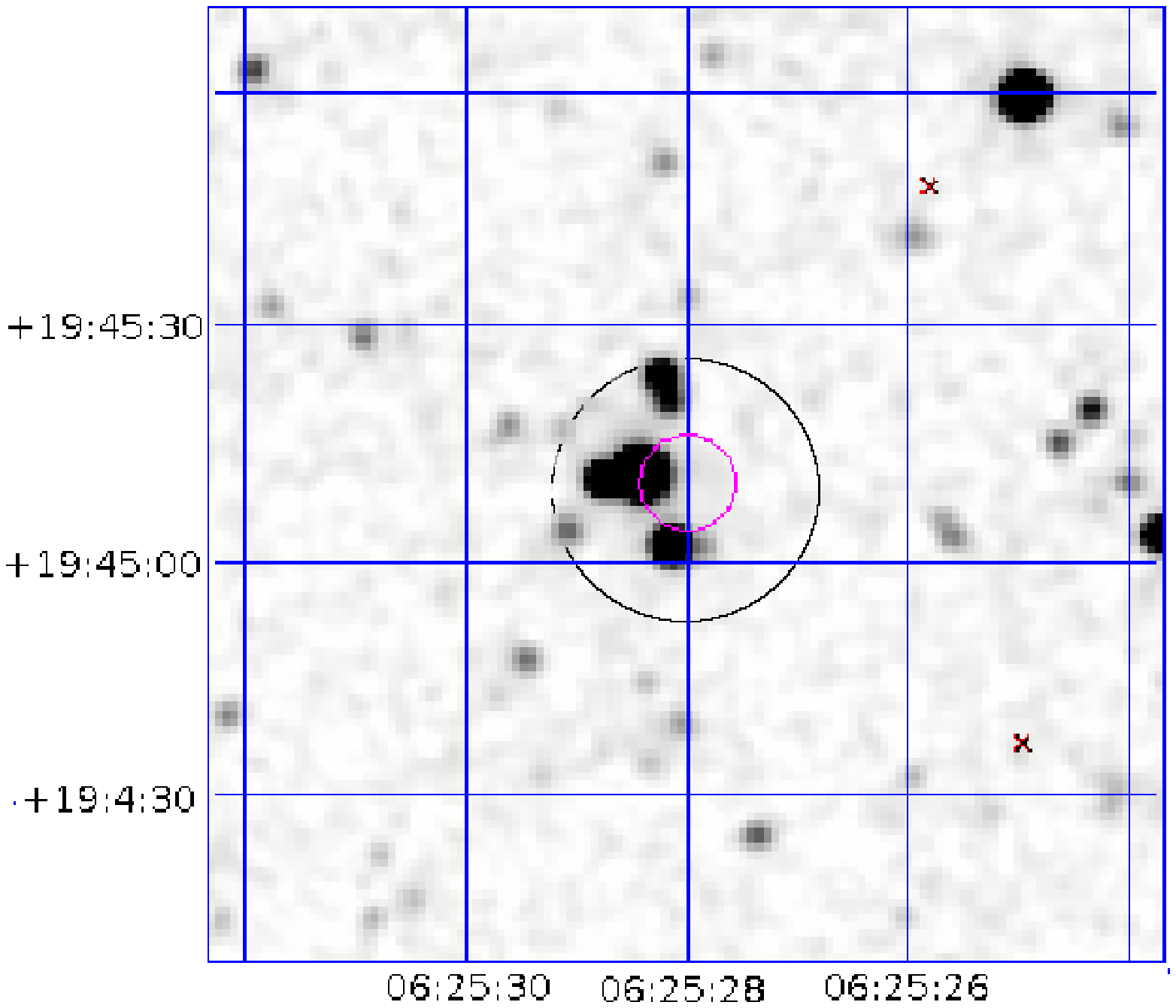}
\end{minipage}\hfill
\begin{minipage}[b]{0.50\linewidth}
\includegraphics[width=\textwidth]{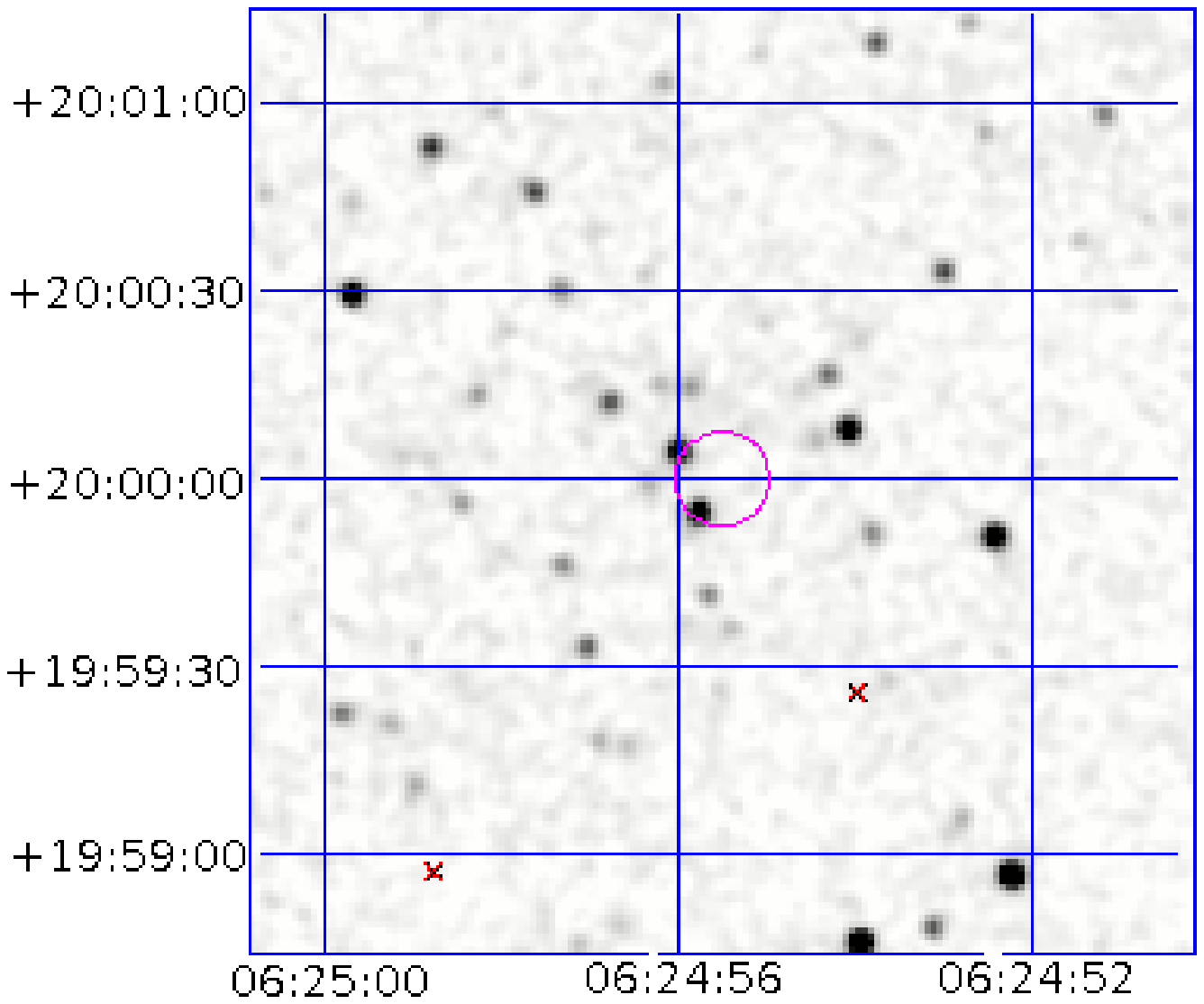}
\end{minipage}\hfill
\begin{minipage}[b]{0.50\linewidth}
\includegraphics[width=\textwidth]{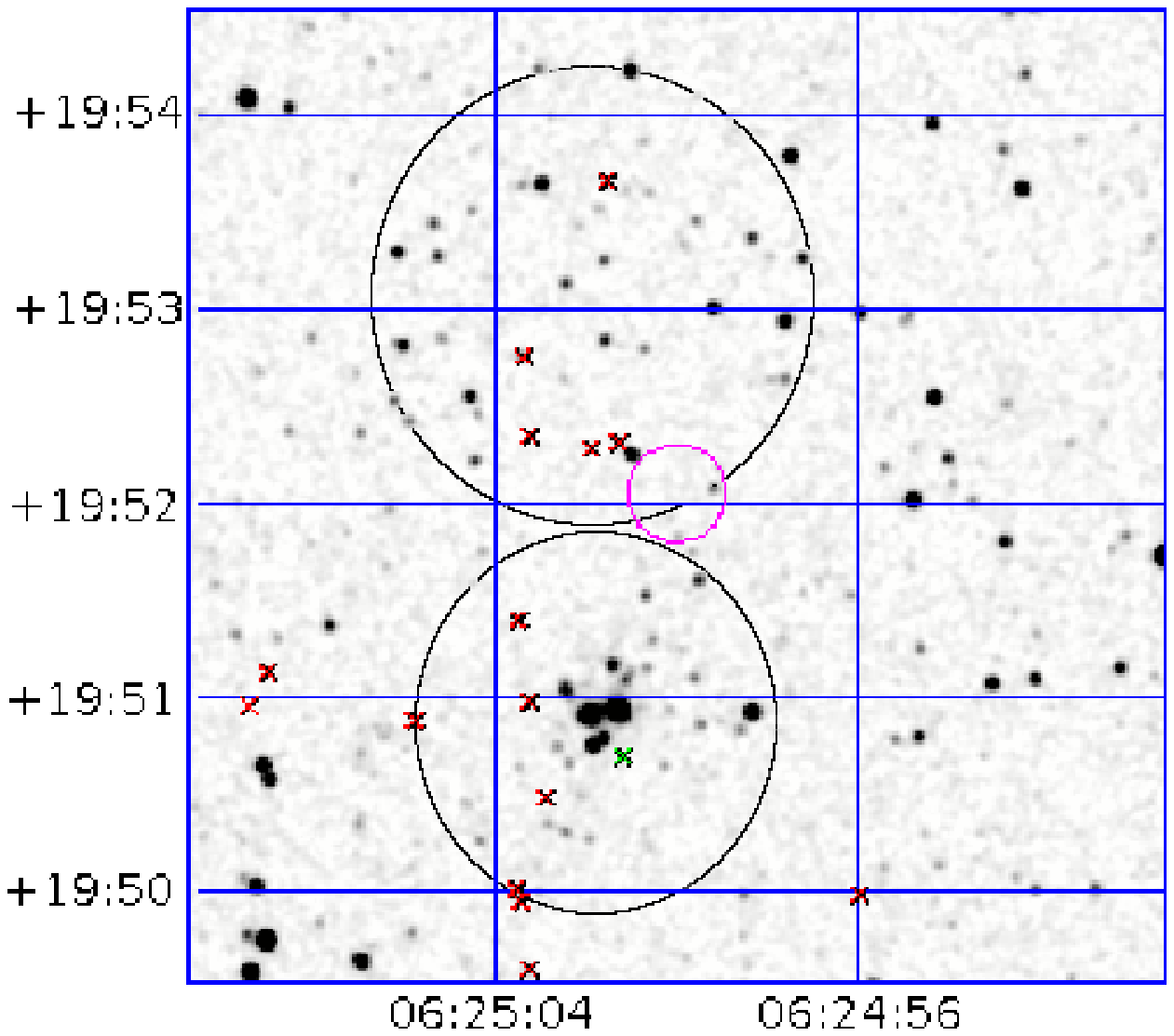}
\end{minipage}\hfill
\caption[]{Top left: $2\arcmin\times2\arcmin$ 2MASS \ks\ image of New\,Cluster\,1; The large
circle corresponds to the RDP radius (Table~\ref{tab5}). Top right: $2\arcmin\times2\arcmin$ 
2MASS \ks\ image of New\,Cluster\,2; the apparent diameter (Table~\ref{tab1}) is shown by the
large circle. Bottom left: $2\arcmin\times2\arcmin$ 2MASS \ks\ image of New\,Cluster\,3. The
RDP radius encompasses the image (Table~\ref{tab5}). Bottom right: $3\arcmin\times3\arcmin$ 
2MASS \ks\ image of the central region of FSR\,911; The large circle to the North corresponds 
to the apparent diameter of the faint clump (Table~\ref{tab1}), and the south one to the
RDP radius (Table~\ref{tab5}). Images provided by the 2MASS Image Service. The small circles 
indicate the central coordinates  (cols.~3 and 4 of Table~\ref{tab1}). Figure orientation: 
North to the top and East to the left. }
\label{fig2}
\end{figure*}

However, when 2MASS (and XDSS) image close-ups were examined (Fig.~\ref{fig2}), we found two very small
compact clusters and an embedded cluster in the area (Table~\ref{tab1}). They have not been reported 
in the literature. New\,Cluster\,1 is located $\approx1\arcmin$ south of the nominal centre of FSR\,911 
(Table~\ref{tab1}), while New\,Cluster\,2 essentially coincides with the central position of Bochum\,1 
(Fig.~\ref{fig1}). The embedded New\,Cluster\,3 is located at the NE edge of the area (Fig.~\ref{fig1}). 
The scattered OB stars that defined Bochum\,1 (\citealt{Moffat75}) are also shown in Fig.~\ref{fig1}. 
Two of these stars are included in New\,Cluster\,1, close to the west border of Bochum\,1. Data for 
Bochum\,1 OB members from \citet{Moffat75} are given in Table~\ref{tab3}, with equatorial coordinates 
and spectral types from SIMBAD\footnote{http://simbad.u-starsbg.fr/simbad/ }. LS46 and LS47 are members 
of New\,Cluster\,1 (Fig.~\ref{fig1}).


New\,Cluster\,3 (Fig.~\ref{fig2}) is looser and appears to be related to the IR source IRAS\,06219+2258. 
New\,Cluster\,1 is the richest one, while New\,Cluster\,2 appears to be more populated than a multiplet 
of stars. Finally, Fig.~\ref{fig2} also shows a blowup of the possible core of the candidate cluster FSR\,911. 
It shows New\,Cluster\,1 to the south and a Clump of faint stars to the north. Cell dimension for counting 
2MASS stars in the FSR algorithm is $3.5\arcmin\times3.5\arcmin$ and the scan step is 20\arcsec\ 
(\citealt{FSRcat}). Their derived core of FSR\,911 must have included contributions of New\,Cluster\,1 and 
the Clump (Fig.~\ref{fig1}). A fundamental question is whether FSR\,911 with its derived large tidal radius 
actually exists as a cluster, association or just an artifact. Another important question is to check 
hierarchically which small structure could be the core of a larger (col.~9 of Table\ref{tab1}) one.

Molecular clouds, particularly giant ones, have multiple cores, part of them actively forming stars
(e.g. \citealt{Yonekura05}). The 3 small clusters in Table~\ref{tab1} might have been related to the
cores of the early molecular cloud associated to Bochum\,1.

\subsection{CMD and structure of Bochum\,1 and companions}
\label{CMD_Bo1}

Photometric properties are investigated by means of CMDs built with 2MASS data in the \jj, \hh\ and \ks\ 
bands. The extractions were performed with VizieR\footnote{\em vizier.u-strasbg.fr/viz-bin/VizieR?-source=II/246}.
We follow the strategy described in \citet{BB07}, and references therein, which is based on photometric
extractions in wide circular fields for statistical representativity of cluster and field stars in terms of 
magnitude and colours. Only stars with errors in \jj, \hh\ and \ks\ smaller than 0.25\,mag (a fraction of about
$75\% - 85\%$ of all stars - \citealt{BB07}) were considered.

We illustrate the procedure by means of the $\jj\times\jh$ and $\jj\times\jk$ CMDs of Bochum\,1 
(Fig.~\ref{fig3}). When compared to the field stars (middle
panels), the CMD extracted from the $R<9\arcmin$ region of Bochum\,1 (top) presents features that
indicate a young age. However, it is evident that contamination by disk stars should be taken into
account before conclusions on the intrinsic CMD morphology of Bochum\,1 are made.

\subsection{Field-star decontamination}
\label{FSD}

Although difficult, field-star decontamination is important to characterise and derive parameters
of star clusters. Several different approaches have been used to this purpose, among them, those of
\citet{Mercer05} and \citet{Carraro06}. The first is based essentially on spatial variations of the
star-count density, but does not take into account colour and magnitude properties. In the latter,
stars of a CMD extracted from an assumed cluster region are subtracted according to colour and
magnitude similarity with the stars of an equal-area comparison field CMD.

In the present case, we apply the statistical algorithm described in \citet{BB07} to quantify the
field-star contamination in the CMDs. The algorithm makes use of both approaches above, in the sense that
relative star-count density together with colour/magnitude similarity between cluster and comparison field
are taken into account simultaneously. It measures the relative number densities of probable
field and cluster stars in cubic CMD cells whose axes correspond to the \jj\ magnitude and the \jh\ and
\jk\ colours. These are the 2MASS colours that provide the maximum variance among CMD sequences for OCs
of different ages (e.g. \citealt{TheoretIsoc}). The algorithm: {\em (i)} divides the full range of
magnitude and colours covered by the CMD into a 3D grid, {\em (ii)} calculates the expected number
density of field stars in each cell based on the number of comparison field stars with similar magnitude
and colours as those in the cell, and {\em (iii)} subtracts the expected number of field stars from each
cell. The algorithm is responsive to local variations of field-star contamination (\citealt{BB07}).
Cell dimensions used here are $\Delta\jj=1.0$, and $\Delta\jh=\Delta\jk=0.25$, which are large enough to
allow sufficient star-count statistics in individual cells and small enough to preserve the morphology of
the CMD evolutionary sequences. For a representative background star-count statistics we use the ring
located within $R_{\rm inf}\le R\le\rx$ around the cluster centre as the comparison field, where
$R_{\rm inf}$ usually represents twice the RDP radius (Sect.~\ref{Bo1_Struc}). We emphasise that
the equal-area field extractions shown in the middle panels of Figs.~\ref{fig3} to \ref{fig5} and \ref{fig9}
serve only for comparisons among the panels. Actually, the decontamination process is carried out with the
large surrounding area as described above.

As extensively discussed in \citet{BB07}, differential reddening between cluster and field stars
is critical for the decontamination algorithm. Large gradients would require large cell sizes
or, in extreme cases, preclude application of the algorithm altogether. Basically, it would be
required, e.g. $|\Delta\jh|\ga\rm cell~size$ (0.25, in the present work) between cluster and comparison
field for the differential reddening to affect the subtraction in a given cell. However, in the present 
cases the CMDs extracted from the cluster region and comparison field (Figs.~\ref{fig3} to \ref{fig5} 
and \ref{fig9}) indicate that the differential reddening is not important.

The decontaminated CMDs are shown in the bottom panels of Figs.~\ref{fig3} to \ref{fig5} and
\ref{fig9}. As expected, most of the contamination is removed, leaving stellar sequences typical 
of young OCs, with a nearly vertical main sequence (MS), and evidence of an important fraction of
pre-MS stars, especially in Bochum\,1 (Fig.~\ref{fig3}) and NGC\,6823 (Fig.~\ref{fig9}). For
illustrative purposes, we provide in Table~\ref{tab2} the full statistics of the decontamination 
process applied to 
the region $R<9\arcmin$ of Bochum\,1, by magnitude bins. Statistically relevant parameters to 
characterise the nature of a star cluster are: 
{\em (i)} \ns\ which, for a given magnitude bin, corresponds to the ratio of the decontaminated number 
of stars to the $\rm1\sigma$ Poisson fluctuation of the number of observed stars, {\em (ii)} \sFS, which 
is related to the probability that the decontaminated stars result from the normal star count fluctuation 
in the comparison field and, {\em (iii)} \fsU, which measures the star-count uniformity of the comparison 
field. Properties of \ns, \sFS, and \fsU, measured in OCs and field fluctuations are discussed in 
\citet{ProbFSR}. Table~\ref{tab2} also provides integrated values of the above parameters, which 
correspond to the full magnitude range spanned by the CMD. The spatial region considered here is that 
sampled by the CMDs shown in the top panels of Fig.~\ref{fig3}.

Star cluster CMDs should have integrated \ns\ values significantly larger than 1 (\citealt{ProbFSR}),
a condition that is fully met by Bochum\,1, with $\ns=9.4$. As a further test of the statistical 
significance of the above results
we investigate star count properties of the field stars. First, the comparison field is divided into 8
sectors around the cluster centre. Next, we compute the parameter
\sFS, which is the $\rm 1\,\sigma$ Poisson fluctuation around the mean of the star counts measured in the 8
sectors of the comparison field (corrected for the different areas of the sectors and cluster extraction). In
a spatially uniform comparison field, \sFS\ is expected to be small. In this context, star clusters should have
the probable number of member stars (\nc) higher than $\sim\sFS$, to minimise the probability that \nc\
arises from fluctuations of a non-uniform comparison field. This condition is fully satisfied, in some cases
reaching the level $\nc\sim4\,\sFS$. Finally, we also provide in Table~\ref{tab2} the parameter \fsU.
For a given magnitude bin we first compute the average number of stars over all sectors $\langle N\rangle$
and the corresponding $\rm 1\sigma$ fluctuation $\sigma_{\langle N\rangle}$; thus, \fsU\ is defined as
$\rm\fsU=\sigma_{\langle N\rangle}/\langle N\rangle$. Non uniformities such as heavy differential reddening
should result in high values of \fsU.

Since we usually work with comparison fields  larger than the possible-cluster extractions, the correction
for the different spatial areas between field and cluster is expected to produce a fractional number
of probable field stars ($n_{fs}^{cell}$) in some cells. Before the cell-by-cell subtraction, the
fractional numbers are rounded off to the nearest integer, but limited to the number of observed stars
in each cell, $n_{sub}^{cell}=NI(n_{fs}^{cell})\leq n_{obs}^{cell}$, where NI represents rounding off to
the nearest integer. The global effect is quantified by means of the difference between the expected number
of field stars in each cell ($n_{fs}^{cell}$) and the actual number of subtracted stars ($n_{sub}^{cell}$).
Summed over all cells, this quantity provides an estimate of the total subtraction efficiency of the
process, \[ f_{sub}(\%)=100\times\sum_{cell}n_{sub}^{cell}/\sum_{cell}n_{fs}^{cell}.\] Ideally, the best
results would be obtained for an efficiency $f_{sub}\approx100\%$. With the assumed grid settings, the
decontamination efficiency turned out to be higher than 96\% in all cases dealt with in this paper.

\begin{table}
\caption[]{Field-star decontamination statistics}
\label{tab2}
\renewcommand{\tabcolsep}{2.3mm}
\renewcommand{\arraystretch}{1.25}
\begin{tabular}{ccccccc}
\hline\hline
$\Delta\jj$&&\multicolumn{5}{c}{Bochum\,1 ($R<9\arcmin$)}\\
\cline{3-7}
 &&\no&\nc&\ns&\sFS&\fsU \\
(mag)&&(stars)&(stars)&&(stars)\\
\cline{1-7}
  8--9&&$4\pm2.0$  &3 &1.5&1.34&0.91   \\
 9--10&&$10\pm3.2$  &4 &1.3&2.00&0.49  \\
10--11&&$11\pm3.3$ &3 &0.9&2.23&0.22 \\
11--12&&$31\pm5.6$ &9 &1.6&4.38&0.21 \\
12--13&&$52\pm7.2$ &10&1.4&3.61&0.07 \\
13--14&&$100\pm10.0$&4&0.4&8.94&0.09  \\
14--15&&$212\pm14.6$&30&2.1&19.385&0.10  \\
15--16&&$499\pm22.3$&144&6.4&38.50&0.10  \\
16--17&&$450\pm21.2$&139&6.6&94.11&0.33  \\
\cline{3-7}
8-17&&$1369\pm37.0$&346&9.4&137.3&0.13\\
\hline
\end{tabular}
\begin{list}{Table Notes.}
\item The upper lines give the statistics in each magnitude bin, while the integrated
values are in the bottom line. See text for details on parameters.
\end{list}
\end{table}

\begin{table}
\caption[]{Bochum\,1 members according to \citet{Moffat75}}
\label{tab3}
\renewcommand{\tabcolsep}{4.2mm}
\renewcommand{\arraystretch}{1.25}
\begin{tabular}{cccc}
\hline\hline
Star&$\alpha(2000)$&$\delta(2000)$&Spectral Type\\
    &(hms)&($^\circ\,\arcmin\,\arcsec$)&\\
(1)&(2)&(3)&(4)\\
\hline
LS\,44 & 06:24:38.4 & $+$19:42:16 & O7.5V \\
LS\,45 & 06:24:55.6 & $+$19:45:55 & B1.5V \\
LS\,46 & 06:25:01.2 & $+$19:50:56 & B \\
LS\,47 & 06:25:01.8 & $+$19:50:54 & B0 \\
LS\,48 & 06:25:04.7 & $+$19:41:30 & G5 \\
LS\,51 & 06:25:24.9 & $+$19:46:00 & B \\
LS\,53 & 06:25:45.5 & $+$19:42:14 & F5 \\
LS\,54 & 06:26:09.0 & $+$19:50:24 & --- \\
\hline
\end{tabular}
\begin{list}{Table Notes.}
\item Coordinates and spectral types from SIMBAD.
\end{list}
\end{table}

\subsection{Fundamental parameters}
\label{FPars}

\begin{figure}
\resizebox{\hsize}{!}{\includegraphics{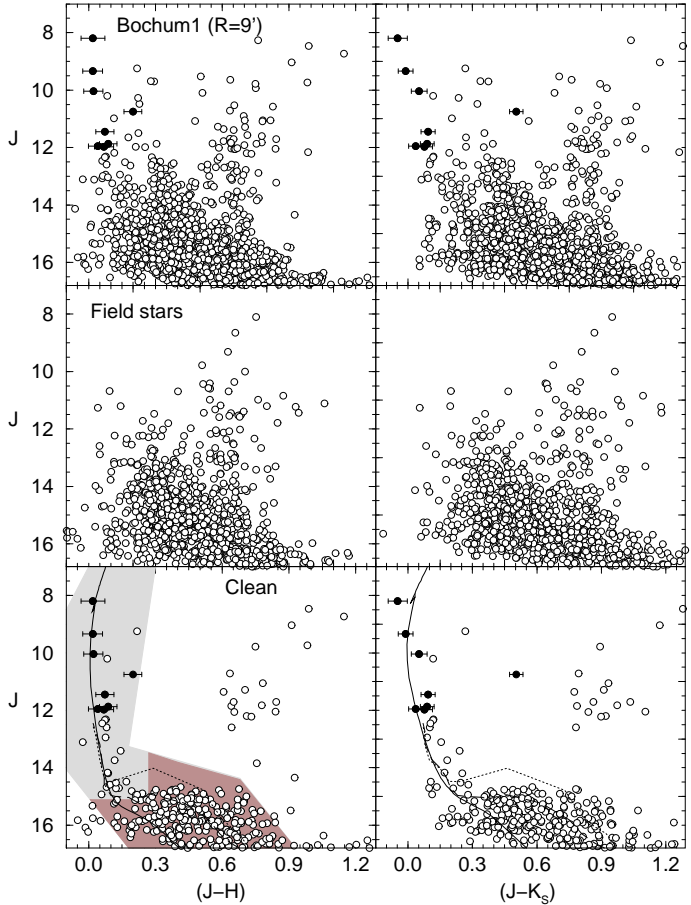}}
\caption{2MASS CMDs extracted from the $R<9\arcmin$ region of Bochum\,1. Top panels: observed photometry
with the $\jj\times\jh$ (left) and $\jj\times\jk$ colours (right). Middle: equal-area comparison field
extracted from the region $49\farcm18 - 50\arcmin$. Bottom panels: decontaminated CMDs showing a
poorly-populated MS and a significant number of PMS stars. Also shown are the 9\,Myr Padova isochrone
(solid line) together with the 5\,Myr (dashed) and 10\,Myr (dotted) PMS tracks (\citealt{Siess2000}).
Light-shaded polygon: Colour-magnitude filter to isolate the MS stars. Heavy-shaded polygon: Colour-magnitude
filter for the PMS stars. Stars with spectral type determined by \citet{Moffat75} are shown as filled
circles. Error bars are not shown to avoid cluttering.}
\label{fig3}
\end{figure}

The young-age features of Bochum\,1 are enhanced in the decontaminated CMDs (bottom panels of
Fig.~\ref{fig3}), especially
a poorly-populated main sequence (MS) and a significant number of pre-main sequence (PMS) stars.
Solar-metallicity isochrones from the
Padova group (\citealt{Girardi02}) computed with the 2MASS filters, and the PMS tracks of \citet{Siess2000}
are used to characterise the age and compute fundamental parameters. Reddening transformations are
$A_J/A_V=0.276$, $A_H/A_V=0.176$, $A_{K_S}/A_V=0.118$, and $A_J=2.76\times\ejh$ (\citealt{DSB2002}), for
a constant $R_V=3.1$, which are based on the extinction curve of \citet{Cardelli89}.

The best-fit was obtained with the 9\,Myr isochrone, apparent distance modulus $\mMJ=13.4\pm0.1$, and
$\ejh=0.11\pm0.02$, which converts to $\ebv=0.35\pm0.06$ and $\aV=1.1\pm0.2$. Considering fit uncertainties,
the age of Bochum\,1 can be set at $9\pm3$\,Myr, consistent with the values of \citet{Moffat75}
and \citet{Yadav03}. The absolute modulus is $\mMo=13.1\pm0.1$, and the distance from the Sun 
$\ds=4.2\pm0.1$\,kpc. Within uncertainties, this value of \ds\ agrees with that of \citet{Moffat75}, 
but it is $\approx50\%$ larger than the value of \citet{Yadav03}.
The 5\,Myr and 10\,Myr PMS tracks set with the above reddening and distance modulus overlap most
of the faint ($\jj\ga15$) and red ($\jh\ga0.3$) stars, which results in an age of $\approx9$\,Myr
for Bochum\,1. This isochrone solution is shown in the bottom panels of Fig.~\ref{fig3}. CMDs in both
colours present comparable results.

The presently derived fundamental parameters are given in Table~\ref{tab4}, where we also provide the
Galactocentric distance (\dgc), which is based on the recently derived value of the Sun's distance to
the Galactic centre $\Rgc=7.2$\,kpc, computed by means of Globular clusters (\citealt{GCProp}).

The remaining objects in the field of Bochum\,1 (Fig.~\ref{fig1}) were analysed in the same way.
Their fundamental parameters are given in Table~\ref{tab4}.

\begin{figure}
\resizebox{\hsize}{!}{\includegraphics{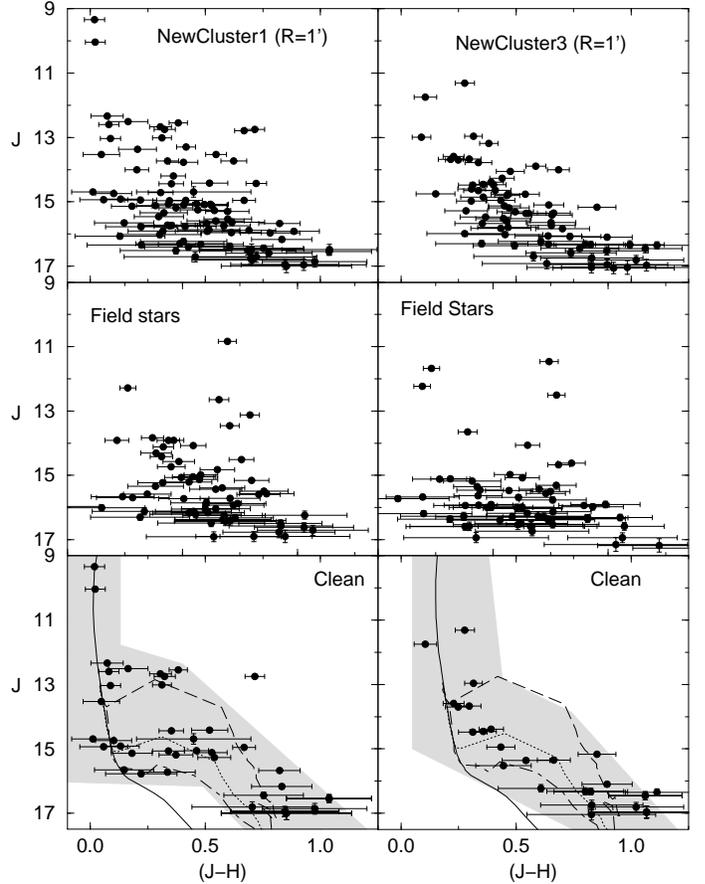}}
\caption{Same as Fig.~\ref{fig3} for the $\jj\times\jh$ CMDs of the $R<1\arcmin$ regions of New\,Cluster\,1
(left panels) and New\,Cluster\,3 (right). Isochrones used for New\,Cluster\,1 are the Padova 7\,Myr (solid
line), and the 1\,Myr (dashed), 5\,Myr (dotted) and 10\,Myr (long-dashed) PMS tracks (\citealt{Siess2000}).
The same isochrone setting was used to characterise the age of New\,Cluster\,3.}
\label{fig4}
\end{figure}

\begin{figure}
\resizebox{\hsize}{!}{\includegraphics{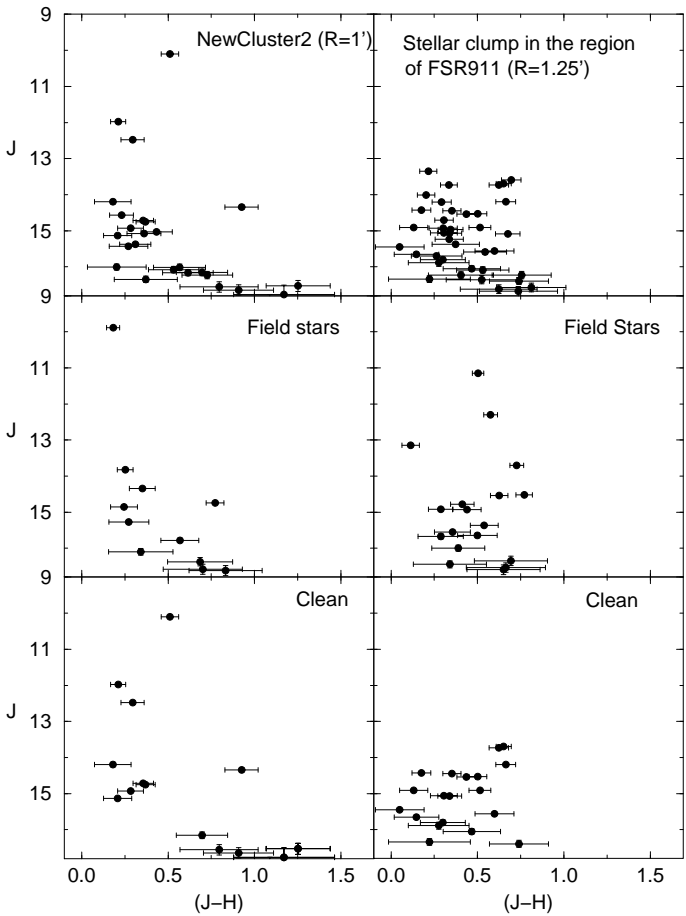}}
\caption{Same as Fig.~\ref{fig3} for the $\jj\times\jh$ CMDs of the $R<1\arcmin$ region of 
New\,Cluster\,2 (left panels) and the $R<1\farcm25$ region of the stellar clump in the region
of FSR\,911 (right). }
\label{fig5}
\end{figure}

Similarly to Bochum\,1, the CMD of New\,Cluster\,1 shows features of a young age (Fig.~\ref{fig4}).
Because of the presence of brighter MS and PMS stars, it appears to be slightly younger than Bochum\,1.
PMS tracks with ages 1\,Myr, 5\,Myr, and 10\,Myr provide a reasonable description of the red and faint 
stellar distribution (bottom-left panel of Fig.~\ref{fig4}). We estimate an age of $7\pm3$\,Myr,
$\mMJ=13.7\pm0.1$, $\ejh=0.15\pm0.01$, $\mMo=13.3\pm0.1$ and $\ds=4.5\pm0.2$\,kpc, which agrees with the 
distance from the Sun of Bochum\,1 (Table~\ref{tab4}).

The poorly-populated decontaminated CMD of New\,Cluster\,3 (bottom-right panel of Fig.~\ref{fig4})
does not provide enough constraints for an independent isochrone fit. In this case we simply use 
the isochrone solution of New\,Cluster\,1 to test whether it is acceptable. Parameters derived in 
this way are $\mMJ=13.9\pm0.2$, $\ejh=0.26\pm0.02$, $\mMo=13.2\pm0.2$ and $\ds=4.3\pm0.4$\,kpc, 
similar to that of Bochum\,1 (Table~\ref{tab4}).

\begin{figure}
\resizebox{\hsize}{!}{\includegraphics{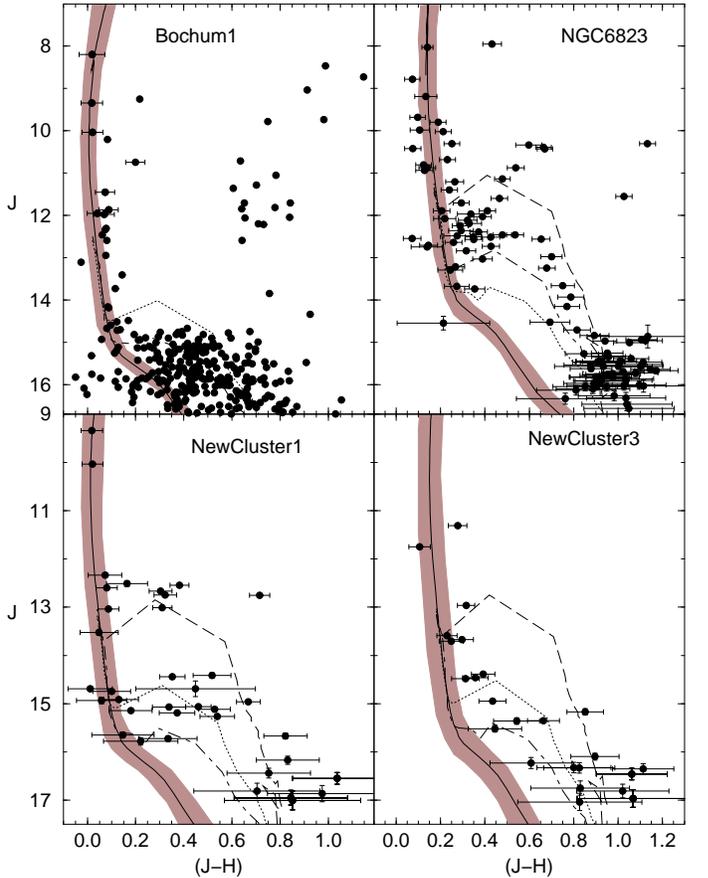}}
\caption{Effects of propagating the $1\sigma$-uncertainties in age, reddening, and distance 
modulus (Table~\ref{tab4}), into the adopted Padova isochrone solution to the MS of Bochum\,1,
NGC\,6823, New\,Cluster\,1 and 3. Field-decontaminated CMDs are used in all cases. Parameter 
variations are enveloped by the shaded polygons.}
\label{fig6}
\end{figure}

\begin{table*}
\caption[]{Fundamental parameters derived in this work}
\label{tab4}
\renewcommand{\tabcolsep}{4.0mm}
\renewcommand{\arraystretch}{1.25}
\begin{tabular}{lccccccc}
\hline\hline
Object&Age&\aV&\ds&\dgc&\xgc&\ygc&\zgc\\
&(Myr)&(mag)&(kpc)&(kpc)&(kpc)&(kpc)&(kpc)\\
(1)&(2)&(3)&(4)&(5)&(6)&(7)&(8)\\
\hline
Bochum\,1      &$9\pm3$&$1.1\pm0.2$&$4.2\pm0.2$&$11.3\pm0.2$&$-11.3\pm0.2$&$-0.88\pm0.02$&$+0.25\pm0.02$  \\
New\,Cluster\,1&$7\pm3$&$1.5\pm0.1$&$4.5\pm0.2$&$11.7\pm0.2$&$-11.6\pm0.2$&$-0.97\pm0.05$&$+0.27\pm0.02$  \\
New\,Cluster\,3$^\dagger$&$7\pm3$&$2.6\pm0.2$&$4.3\pm0.4$&$11.5\pm0.4$&$-11.5\pm0.4$&$-0.91\pm0.09$&$+0.26\pm0.02$ \\
NGC\,6823&$4\pm2$&$2.7\pm0.1$&$2.0\pm0.1$&$6.4\pm0.1$&$-6.2\pm0.2$&$+1.76\pm0.04$&$0.00\pm0.01$  \\
\hline
\end{tabular}
\begin{list}{Table Notes.}
\item Col.~2: Age derived in this paper with 2MASS data. Col.~3: $\rm\aV=3.1\,\ebv$. Col.~4: Distance 
from the Sun. Col.~5: Galactocentric distance calculated with $\Rgc=7.2$\,kpc (\citealt{GCProp}) as the
distance of the Sun to the Galactic centre. Cols.~6-8: Positional components with respect to the Galactic 
plane. $(\dagger)$: Parameters computed for the same isochrone solution as New\,Cluster\,1.
\end{list}
\end{table*}

Finally, in Fig.~\ref{fig6} we examine how the isochrone fit uncertainties affect the
quality of the adopted solution. It shows, for the 4 cases with fundamental parameters and 
errors derived in the present work (Table~\ref{tab4}), the adopted Padova isochrone solution 
to the MS together with the solutions produced by the $1\sigma$-variations applied to the 
adopted age, reddening and distance modulus. With 3 independent parameters and 3 different
values each, a total of 27 solutions were taken into account. Thus, for clarity we show 
in Fig.~\ref{fig6} the {\em best-fit} together with the envelope of solutions. Note that 
NGC\,6823 is analysed in Sect.~\ref{CMD_N6823}. Since the objects shown in Fig.~\ref{fig6} 
are very young, most of their stellar content, especially the red ones, should correspond 
to the PMS phase. Indeed, Fig.~\ref{fig6} also shows that the colour range spanned by these 
red stars cannot be accounted for by uncertainties associated to the MS isochrone fit.

\subsection{Cluster structure}
\label{Bo1_Struc}

We examine the structure of the above objects by means of the projected radial distribution of
the number density of stars around the centre. We work with RDPs built with colour-magnitude filtered
(bottom panels of Figs.~\ref{fig3}, \ref{fig4}, and \ref{fig9}) photometry, which minimises contamination of
non-cluster stars and produces more intrinsic profiles (e.g. \citealt{BB07} and references therein).
To describe the RDPs we use the analytical profile $\sigma(R)=\sigma_{bg}+\sigma_0/(1+(R/R_C)^2)$,
where $\sigma_{bg}$ is the residual background density, $\sigma_0$ is the central density of stars,
and \rc\ is the core radius. This function is similar to the \cite{King1962} profile usually applied
to the central parts of globular clusters.

\begin{figure}
\resizebox{\hsize}{!}{\includegraphics{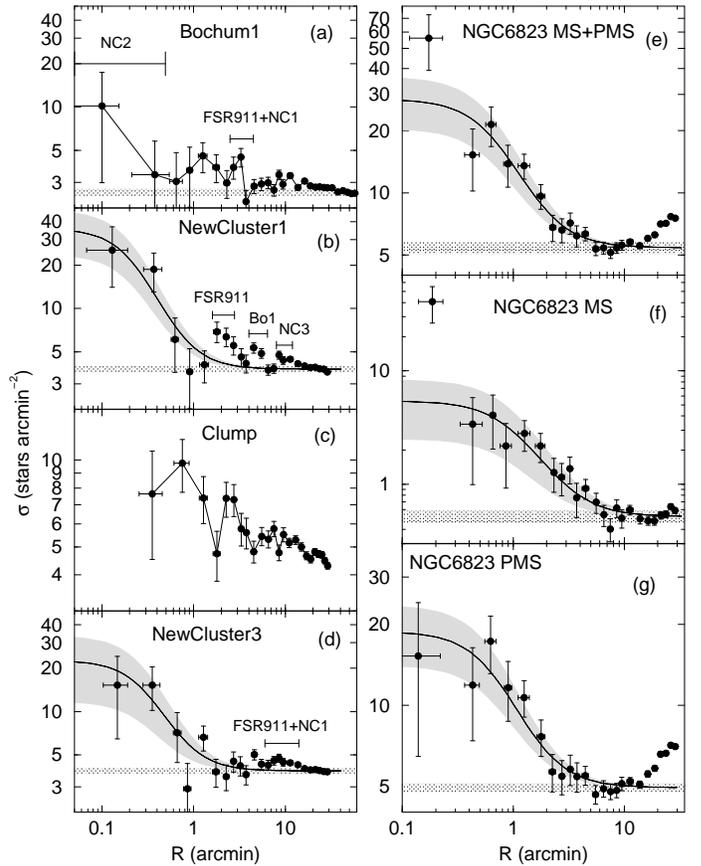}}
\caption{Stellar RDPs (filled circles) built with colour-magnitude photometry. The best-fit King-like
profile is shown as a solid line in panels (b), (d) - (g), where the background level (horizontal shaded
polygon) and the $1\sigma$ King fit uncertainty (gray region) are also shown. The RDPs of Bochum\,1 (a)
and the stellar clump in the region of FSR\,911 (c) cannot be fitted. RDPs built with the MS (f) and PMS
(g) stars of NGC\,6823 are shown separately. Angular scale is used.}
\label{fig7}
\end{figure}

The RDPs of Bochum\,1, New\,Cluster\,1, and New\,Cluster\,3 are shown in Fig.~\ref{fig7} (left panels). 
Bochum\,1, in particular, has a low-contrast profile disturbed by the presence of New\,Cluster\,1 and 
the possible presence of FSR\,911. As a result, this irregular RDP cannot be fitted with the adopted 
King-like profile. New\,Cluster\,2 fits right into the central bin of the RDP of Bochum\,1 (Fig.~\ref{fig7}, 
panel a). Despite the irregularities, this RDP still presents a density gradient decreasing for
larger radii.

New\,Cluster\,1 and New\,Cluster\,3, on the other hand, have profiles that, despite 
bumps due to neighbouring objects, follow the King-like function, especially the former. To arrive
at the fits shown in Fig.~\ref{fig7}, the RDP points that correspond to other objects were excluded.
We also estimate the cluster RDP radius \rl, which corresponds to the distance from the cluster
centre where RDP and background become statistically indistinguishable (e.g. \citealt{DetAnalOCs}).
For the purposes of the present work, we adopt \rl\ as cluster size.

\begin{table*}
\caption[]{Structural parameters measured in the RDPs built with colour-magnitude filtered photometry}
\label{tab5}
\renewcommand{\tabcolsep}{2.0mm}
\renewcommand{\arraystretch}{1.4}
\begin{tabular}{lcccccccccc}
\hline\hline
&&&&\multicolumn{6}{c}{RDP}\\
\cline{5-10}
Cluster&$1\arcmin$&&$\sigma_{bg}$&$\sigma_0$&$\delta_c$&\rc&\rl&\rc&\rl \\
       &(pc)&&$\rm(stars\,arcmin^{-2})$&$\rm(stars\,arcmin^{-2})$&&(\arcmin)&(\arcmin)&(pc)&(pc)\\
(1)&(2)&&(3)&(4)&(5)&(6)&(7)&(8)&(9)\\
\hline
Bochum\,1      &1.309&&$2.6\pm0.1$&---&---&--- &$18\pm2$ &---&$23\pm3$\\
New\,Cluster\,1&1.317&&$3.8\pm0.1$&$32\pm12$&$9.4\pm3.1$&$0.23\pm0.08$ &$1.0\pm0.1$ &$0.30\pm0.10$&$1.3\pm0.1$\\
New\,Cluster\,3$^\dagger$&1.256&&$3.9\pm0.1$&$18.7\pm10.8$&$5.8\pm2.8$&$0.31\pm0.15$ &$2.0\pm0.2$&$0.39\pm0.19$&$2.5\pm0.3$\\
NGC\,6823$^a$  &0.593&&$5.4\pm0.1$&$23.0\pm7.9$&$4.2\pm0.3$&$0.74\pm0.19$ &$5.0\pm0.5$ &$0.43\pm0.11$&$3.0\pm0.3$\\
NGC\,6823$^b$    &0.593&&$0.5\pm0.1$&$4.9\pm2.9$&$10.4\pm3.2$&$0.97\pm0.43$ &$7.0\pm1.0$ &$0.57\pm0.25$&$4.1\pm0.6$\\
NGC\,6823$^c$     &0.593&&$4.9\pm0.1$&$13.9\pm4.7$&$3.8\pm0.9$&$0.77\pm0.21$ &$5.0\pm1.0$ &$0.46\pm0.12$&$3.0\pm0.6$\\
\hline
\end{tabular}
\begin{list}{Table Notes.}
\item Col.~2: arcmin to parsec scale. To minimise degrees of freedom in RDP fits with the King-like
profile (see text), $\sigma_{bg}$ was kept fixed (measured in the respective comparison fields) while
$\sigma_0$ and \rc\ were allowed to vary. Col.~5: cluster/background density contrast
($\delta_c=1+\sigma_0/\sigma_{bg}$), measured in colour-magnitude filtered RDPs. $(\dagger)$: Absolute
values computed for the same isochrone solution of New\,Cluster\,1. (a): Measured in the RDP that includes
MS and PMS stars. (b): MS stars only. (c): PMS stars only.
\end{list}
\end{table*}

The structural parameters derived as described above are given in Table~\ref{tab5}, where we also
include the density contrast parameter $\delta_c=1+\sigma_0/\sigma_{bg}$. As expected from the RDPs
in Fig.~\ref{fig7}, Bochum\,1 presents the lowest contrast profile among the objects in that region.

\section{NGC\,6823: a nearby template of a young open cluster}
\label{N6823}

NGC\,6823 has been studied by \citet{Barkhatova57} and more recently by \citet{Kharchenko05}. 
The latter authors derive the age of 10\,Myr, $\ebv=0.84$, $\ds=1.9$\,kpc, $\rc=4\farcm2$, and
a cluster radius $\rm R_{clus}=16\farcm2$. WEBDA provides $\ebv=0.85$, $\ds=1.9$\,kpc and an
age of 6\,Myr. We use NGC\,6823 as a relatively nearby template OC. It has a prominent
core (\citealt{Pigu00}), and the small embedded cluster Cr\,404 is projected not far from NGC\,6823. 
NGC\,6823 and Cr\,404 are part of the association Vul\,OB1 with a diameter 
of $170\arcmin\times130\arcmin$. \citet{Massey95} studied NGC\,6823/Vul\,OB1 deriving an average 
reddening $\ebv=0.89$, a distance $\ds=2.3$\,kpc and  ages in the range 2-7\,Myr. The H\,II region
Sh2-86 (\citealt{Sharpless59}) with an angular diameter of 40\arcmin\ is included in Vul\,OB1. The 
kinematic distance of Sh2-86 is 1.9\,kpc (\citealt{BrBl93}).
 
\begin{figure}
\resizebox{\hsize}{!}{\includegraphics{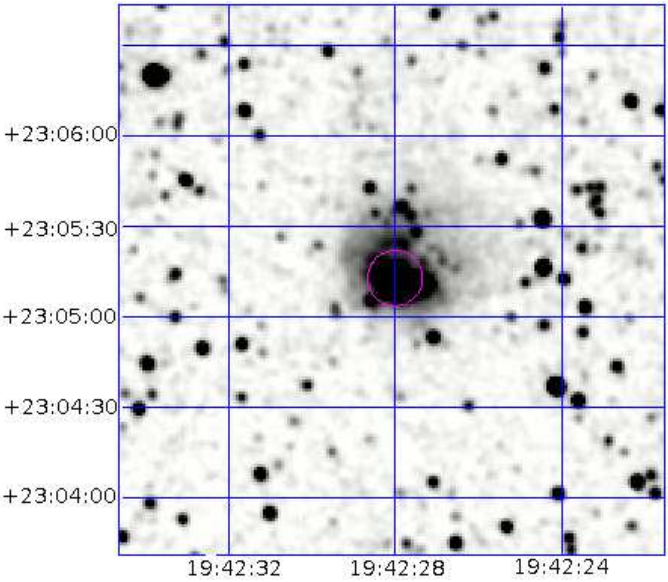}}
\caption[]{$3\arcmin\times3\arcmin$ 2MASS \ks\ image of the embedded cluster Cr\,404 in the nebula 
NGC\,6820. Figure orientation: North to the top and East to the left.}
\label{fig8}
\end{figure}

Cr\,404 was first recognised as a star cluster by \citet{Collinder31}. It is embedded in the small
angular size nebula NGC\,6820. In modern classifications it is a typical embedded cluster 
(\citealt{Hodapp94}). It is projected just outside NGC\,6823. A 2MASS \ks\ image of the embedded 
cluster Cr\,404 in the nebula NGC\,6820 is shown in Fig.~\ref{fig8}. We conclude that the Vul\,OB1 
complex shows an OC with a prominent core and an embedded cluster with hardly any trace of a halo. It 
is extremely compact and unresolved by 2MASS photometry. Cr\,404 requires a large telescope for a deeper 
analysis.

\subsection{CMD and structure of NGC\,6823}
\label{CMD_N6823}

\begin{figure}
\resizebox{\hsize}{!}{\includegraphics{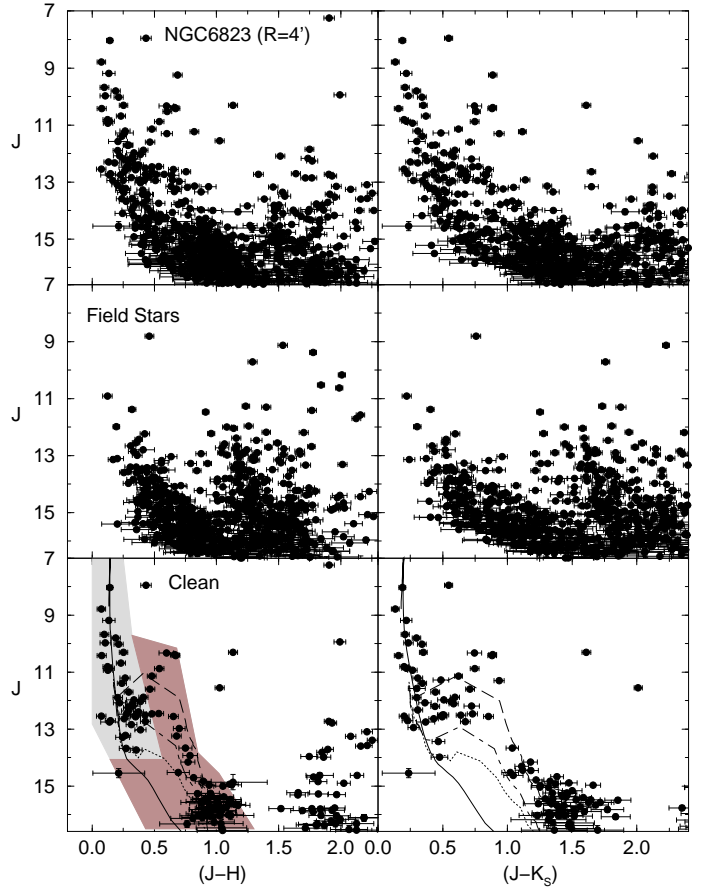}}
\caption{Same as Fig.~\ref{fig3} for the $R<4\arcmin$ region of NGC\,6823. The isochrones 
(bottom-left) used are the 4\,Myr (Padova) and the PMS tracks of 1\,Myr (dashed line), 5\,Myr 
(dot-dashed), and 10\,Myr (dotted). Light-shaded polygon: colour-magnitude filter to isolate
the MS stars. Heavy-shaded polygon: colour-magnitude filter for the PMS stars.}
\label{fig9}
\end{figure}

We apply to NGC\,6823 the same analysis as for Bochum\,1 (Sect.~\ref{CMD_Bo1}). The $\jj\times\jh$ 
CMD extracted from the $R<3\arcmin$ of NGC\,6823 is shown in Fig.~\ref{fig9}. A conspicuous, 
nearly-vertical MS can be seen in the decontaminated CMD (bottom-left panel), together with a
group of faint and red PMS stars. Such CMD morphology can be well described with a 4\,Myr
Padova isochrone and the 1\,Myr, 5\,Myr, and 10\,Myr PMS tracks. Acceptable fits are obtained 
with ages in the range 2-7\,Myr. The fit in Fig.~\ref{fig9} was obtained for $\mMJ=12.3\pm0.1$, 
$\ejh=0.27\pm0.01$,which converts to $\ebv=0.86\pm0.06$ and $\aV=2.7\pm0.2$. The absolute modulus 
is $\mMo=11.5\pm0.1$, and the distance from the Sun is $\ds=2.0\pm0.1$\,kpc (Table~\ref{tab4}).
These values agree with those in \citet{Kharchenko05} and WEBDA.


The RDPs of NGC\,6823 are shown in Fig.~\ref{fig7}. Besides the RDP that includes MS and PMS stars,
we also consider both stellar distributions separately. The King-like function describes well most
of the profiles, except for the innermost radial bin in the MS and MS$+$PMS RDPs, which indicates
an excess of MS stars near the cluster centre ($R\la0\farcm2\arcmin\approx0.1$\,pc). This feature 
is characteristic of post-core collapse globular clusters (\citealt{Trager95}). We note that a
post-core collapse feature in the RDP of OCs has been previously detected, for instance, in the 
$\sim1$\,Gyr old cluster NGC\,3960 (\citealt{BB06}). However, at the young age of NGC\,6823, a
possible explanation for the central stellar density is an enhanced fragmentation in the central
parts of the parent molecular cloud, and/or primordial dynamical evolution. Another conspicuous 
feature is the excess of PMS stars for $R\ga10\arcmin$.

\section{Mass functions and stellar content}
\label{MF}

\begin{figure}
\resizebox{\hsize}{!}{\includegraphics{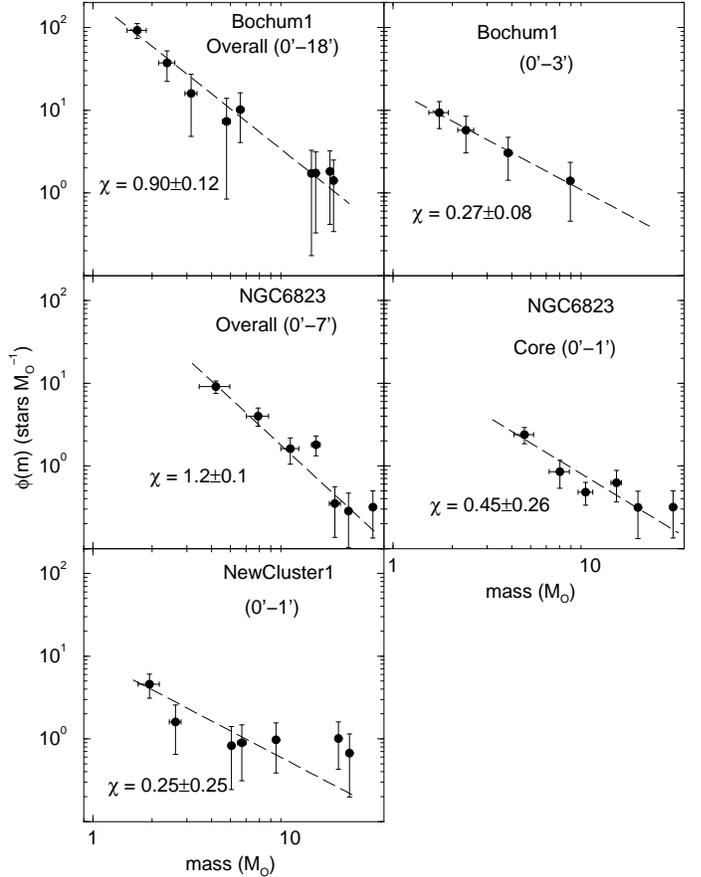}}
\caption{Mass functions of Bochum\,1 (top panels), NGC\,6823 (middle), and New\,Cluster\,1
(bottom). The left panels show the MF derived for the whole cluster region, while the right ones
contain the MF for the inner region, the core in the case of NGC\,6823. The dashed line shows
the mass function fit.}
\label{fig10}
\end{figure}

For a deeper analysis of the stellar distribution we build mass functions (MFs) for the MS stars with
the \jj, \hh, and \ks\ bands independently (see, e.g. \citealt{DetAnalOCs}). The MFs of Bochum\,1,
New\,Cluster\,1, and NGC\,6823 are built with the stars isolated with the respective colour-magnitude
filters (Figs.~\ref{fig3}, \ref{fig4}, and \ref{fig9}), which minimise contamination by
field stars (e.g. \citealt{BB07}). In all cases we consider the full radial extent of the
objects and inner regions, the core in the case of NGC\,6823. The MFs are shown in
Fig.~\ref{fig10}. The MS mass ranges are $1.5-17\,\ms$ for Bochum\,1, $3.5-27\,\ms$ for NGC\,6823,
and $1.3-21\,\ms$ for New\,Cluster\,1.

The MFs are well described by the function $\phi(m)=\phi_0\,m^{-(1+\chi)}$. With $\chi\approx0.9$,
the overall MF of Bochum\,1 (top panels) is somewhat flatter than the slope $\chi=1.35$ of \citet{Salpeter55}
initial mass function (IMF). However, the MF extracted for an inner region ($R<3\arcmin$) shows an even
flatter slope, $\chi\approx0.3$. Such flat values, especially the inner one, reflect large-scale
mass segregation in this young object, which suggests primordial dynamical evolution and/or 
star-forming effects (e.g. \citealt{DetAnalOCs}).

A similar picture applies to NGC\,6823 (middle panels), although in this case the overall MF slope 
approaches the Salpeter one. We note that the present overall slope is steeper than that derived by
\citet{Massey95}, $\chi=0.3$. New\,Cluster\,1 also presents a flat MF slope, $\chi\approx0.25$
(bottom).

In Table~\ref{tab6} we quantify the stellar content for Bochum\,1, NGC\,6823, and
New\,Cluster\,1. We consider MS and PMS stars separately (for simplicity we assume a canonical
mass of 1\,\ms\ for the PMS stars). We estimate that only $\approx23\%$ of the stars in Bochum\,1
have already reached the MS. MS and PMS stars taken together, the mass of Bochum\,1 is
$\approx720\,\ms$. In NGC\,6823 the fractions of MS and PMS stars are more evenly distributed.
It is more massive than Bochum\,1, with a mass of about $1150\,\ms$. New\,Cluster\,1 has a
small number of MS and PMS stars ($\approx20$), which results in a very-low mass of $\approx74\,\ms$.

Crowding and completeness should not be important for the above arguments, because these objects
are mostly poorly-populated and sparse, and besides, faint MS stars are not included 
in the colour-magnitude filters (Figs.~\ref{fig3}, \ref{fig4}, and \ref{fig9}).

\begin{table*}
\caption[]{Stellar content and mass functions}
\label{tab6}
\renewcommand{\tabcolsep}{2.25mm}
\renewcommand{\arraystretch}{1.25}
\begin{tabular}{cccccccccccc}
\hline\hline
&\multicolumn{5}{c}{MS}&&\multicolumn{2}{c}{PMS}&&\multicolumn{2}{c}{MS$+$PMS}\\
\cline{2-6}\cline{8-9}\cline{11-12}
$\Delta\,R$&$\Delta\,m_{MS}$&N&M&$\phi_0$&$\chi$&&N&M&&N&M\\
(\arcmin)&(\ms)&(stars)&(\ms)&$(\rm stars\,\ms^{-1})$&&&(stars)&(\ms)&&(stars)&(\ms)\\
(1)&(2)&(3)&(4)&(5)&(6)&&(7)&(8)&&(9)&(10)\\
\hline
\multicolumn{12}{c}{Bochum\,1 - Age $=9\pm3$\,Myr}\\
\hline
0---3&1.5---8.2&$16\pm2$&$46\pm6$&$17.8\pm1.5$&$0.27\pm0.08$&&$15\pm8$&$15\pm8$&&$31\pm9$&$61\pm10$\\
0---18&1.5---17&$128\pm11$&$323\pm29$&$220\pm32$&$0.90\pm0.12$&&$397\pm49$&$397\pm49$&&$525\pm55$&$720\pm60$\\
\hline
\multicolumn{12}{c}{NGC\,6823 - Age $=4\pm2$\,Myr}\\
\hline
0---1&4.1---27&$15\pm2$&$158\pm23$&$19.5\pm9.7$&$0.45\pm0.26$&&$29\pm5$&$29\pm5$&&$44\pm6$&$177\pm25$\\
0---7&3.5---27&$65\pm4$&$1046\pm79$&$224\pm49$&$1.19\pm0.11$&&$92\pm20$&$92\pm20$&&$157\pm22$&$1150\pm85$\\
\hline
\multicolumn{12}{c}{New\,Cluster\,1 - Age $=7\pm3$\,Myr}\\
\hline
0---1&1.3---21&$10\pm1$&$64\pm11$&$9.3\pm3.3$&$0.25\pm0.25$&&$10\pm4$&$10\pm4$&&$20\pm4$&$74\pm12$\\
\hline
\end{tabular}
\begin{list}{Table Notes.}
\item Col.~1: Spatial region where the MF is computed. Col.~2: MS mass range. Cols.~3-6: Stellar
content and MF of the MS stars. Cols.~7-8: PMS stellar content. Cols.~9-10: Total stellar content.
\end{list}
\end{table*}

\begin{figure}
\resizebox{\hsize}{!}{\includegraphics{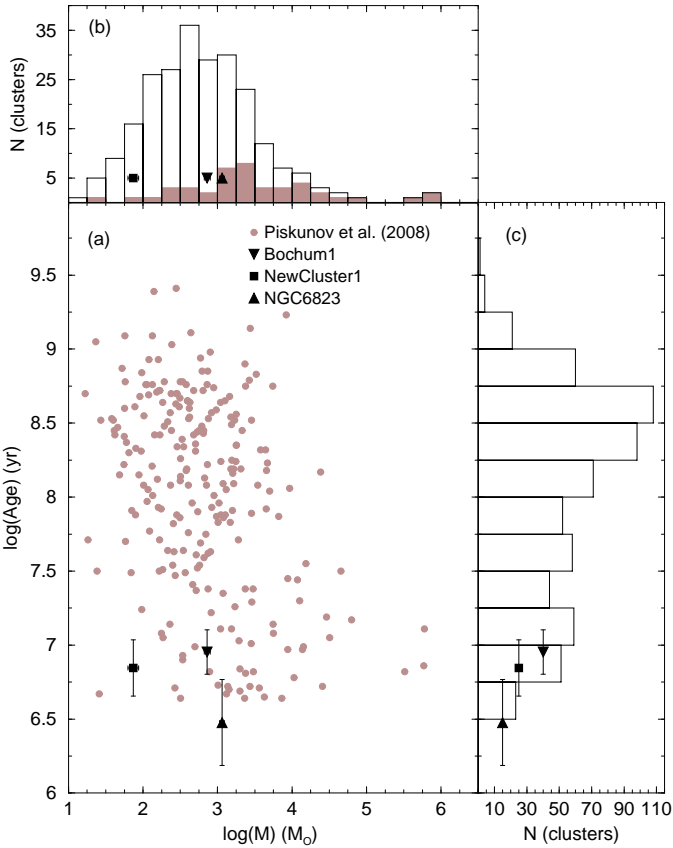}}
\caption{The age and mass of Bochum\,1, New\,Cluster\,1, and NGC\,6823 are compared to the
set of parameters derived by \citet{Piskunov08} for a sample of nearby OCs (panel a). Panels
(b) and (c): mass and age distributions, respectively. The shaded histogram in (b) corresponds 
to clusters younger than 20\,Myr.}
\label{fig11}
\end{figure}

Finally, in Fig.~\ref{fig11} we compare the presently measured age and mass values for Bochum\,1, 
New\,Cluster\,1, and NGC\,6823 with those derived by \citet{Piskunov08}\footnote{Data taken from the
VizieR On-line Data Catalog: J/A+A/477/165} for a relatively large sample
of nearby OCs. Our mass value for NGC\,6823 (Table~\ref{tab6}) is about half that estimated by 
\citet{Piskunov08} which, considering the different methods, can be taken as a reasonable agreement.
Based on the mass distribution of the clusters with any age (panel b), the genuine young OC NGC\,6823 
is more massive than the average cluster mass. A similar conclusion applies to Bochum\,1. On the other 
hand, New\,Cluster\,1 is among the least massive dynamical survivors of the early phases. However, when 
only clusters younger than 20\,Myr are considered, Bochum\,1, and especially New\,Cluster\,1, populate 
the low-mass tail of the \citet{Piskunov08} distribution. It is clear that we are dealing with clusters 
or stellar groups that survived the infanticide phase of embedded clusters (\citealt{LL2003}). 

\section{Discussion}
\label{Disc}

In this section we compare the properties of the present objects to those of
a collection of well-studied OCs analysed by our group following similar methods.

\subsection{Diagnostic diagrams}
\label{DD}

We further investigate the nature of Bochum\,1 (and the other objects in the area) with diagrams that
deal with relations among astrophysical parameters of OCs in different environments, which have been
introduced by \citet{DetAnalOCs}. As reference we use a sample of bright nearby OCs (\citealt{DetAnalOCs};
\citealt{N4755}), and a group of OCs projected against the central parts of the Galaxy (\citealt{BB07}).
Also included is the young ($\sim1.3$\,Myr) OC NGC\,6611 (\citealt{N6611}) for comparison with a gravitationally
bound object of similar age. These OCs have ages in the range $\sim1.3$\,Myr to $\sim7$\,Gyr, and Galactocentric
distances within $\rm5.8\la\dgc(kpc)\la8.1$.

Panels (a) and (b) examine the dependence of cluster (\rl) and core (\rc) radii on cluster age, respectively.
While New\,Cluster\,3 and NGC\,6823 have \rl\ similar to that of NGC\,6611 (but apparently smaller than the 
radii of young OCs), Bochum\,1 and New\,Cluster\,1 appear to be abnormally large and small, respectively. A 
similar relation occurs for the core radii (note that it was not possible to derive \rc\ for Bochum\,1).  
Most of the small-radii OCs (especially in \rl) occurs at $\sim0.5-1$\,Gyr, the typical time-scale of OC 
disruption processes near the Solar circle (e.g. \citealt{Bergond2001}; \citealt{Lamers05}).

Core and \rl\ of the OCs in the reference sample follow the relation $\rl=(8.9\pm0.3)\times
R_{\rm C}^{(1.0\pm0.1)}$ (panel c). Similar relations between core and RDP radii were also found
by \citet{Nilakshi02}, \citet{Sharma06}, and \citet{MacNie07}.Within uncertainties, NGC\,6823, 
New\,Cluster\,3 (and NGC\,6611) fit in that relation. The deviant case is again New\,Cluster\,1. Dependence 
of OC size on Galactocentric distance is suggested by panel (d), as previously discussed by \citet{Lynga82} 
and \citet{Tad2002}. Except for New\,Cluster\,1 and New\,Cluster\,3, the remaining objects follow the trend.

\begin{figure}
\resizebox{\hsize}{!}{\includegraphics{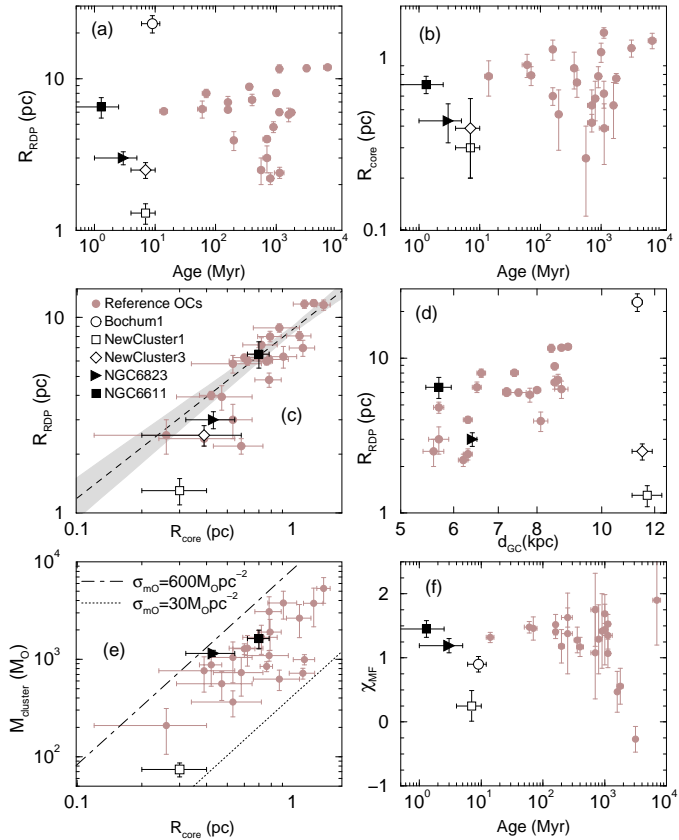}}
\caption{Relations involving astrophysical parameters of open clusters. Gray-shaded circles:
reference OCs. Core radius and cluster (\rl) size (c) are related by $\rl\approx8.9\rc$. Core 
radius and cluster mass (e) follow the relation $\rm M_{clus}\approx13.8\sigma_{M0}\,\rc^2$.}
\label{fig12}
\end{figure}

When the mass radial distribution follows a King-like profile (e.g. \citealt{OldOCs};
\citealt{StrucSCs}; \citealt{PNOCs}), the cluster mass inside \rl\ can be computed as a function
of the core radius (\rc) and the central mass-surface density ($\sigma_{M0}$), $\rm
M_{clus}=\pi\,R^2_{C}\sigma_{M0}\ln\left[1+\left(\rl/\rc\right)^2\right]$. With the above 
relation (panel c) between \rc\ and \rl, this equation becomes $\rm M_{clus}\approx13.8\sigma_{M0}\,R^2_{C}$.
The observed relation of core radius and cluster mass is examined in panel (e). The reference 
OCs, together with NGC\,6823, New\,Cluster\,1, and NGC\,6611 are contained within King-like 
mass-distributions with central densities within $\rm30\la\sigma_{M0}\,(\ms\,pc^{-2})\la600$. 
Similarly to the central number-density (Table~\ref{tab5}), New\,Cluster\,1 presents one of 
the lowest central mass densities among the OCs included in Fig.~\ref{fig12}.

When the MF slope is considered, Bochum\,1 and especially New\,Cluster\,1, appear to 
have MFs flatter than those of similarly young OCs (panel f). On the other hand, their slopes 
are equivalent to those derived for some of the old OCs in the reference sample. In most cases, 
flat MFs reflect advanced dynamical evolution (e.g. \citealt{DetAnalOCs}).

\begin{figure}
\resizebox{\hsize}{!}{\includegraphics{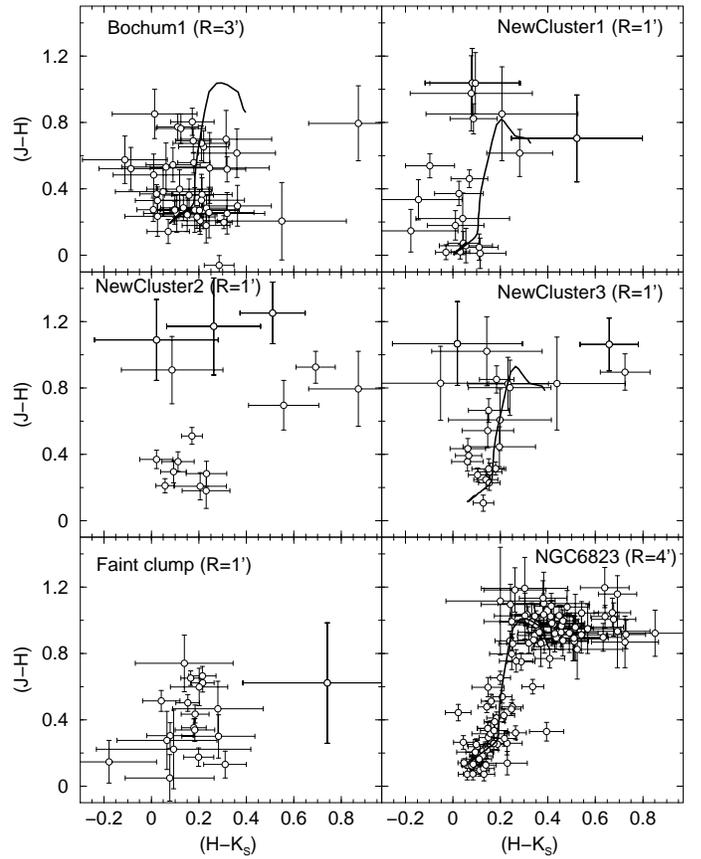}}
\caption{Field-decontaminated colour-colour diagrams of the objects. The respective Padova 
isochrones are shown as reference for the MS stars. }
\label{fig13}
\end{figure}

We conclude that New\,Cluster\,2 is the core of Bochum\,1, while New\,Cluster\,1 could be the
core of FSR\,911, if the latter is confirmed as a stellar system.

\subsection{Field-decontaminated colour-colour diagrams}
\label{2CD}

We show in Fig.~\ref{fig13} the $(\hh-\ks)\times\jh$ colour-colour diagrams for the
present objects, built with decontaminated photometry to minimise noise. The template NGC\,6823
presents well-populated MS and PMS ($\jh\ga0.8$) sequences. Bochum\,1 and its neighbours are
older than NGC\,6823, as denoted by the depletion of PMS stars (\citealt{N4755}). The
faint-clump diagram is consistent with being part of Bochum\,1 population surroundings. Since its
RDP decays radially, it is an overdensity, which does not exclude the possibility of the clump
being a fragment of the Bochum\,1 association, or of FSR\,911.

Low-mass embedded clusters in general do not harbour any ionising star (\citealt{Soares05}), 
which appears to be the case of New Cluster\,3. They lose important fractions of the primordial 
mass (\citealt{GoBa06}), become gravitationally unstable, and are expected to dissolve in a 
timescale of a few Myr (\citealt{Soares08}; \citealt{GoBa06}).

Compact young clusters (CYCs) appear to have survived the gas mass loss during the embedded cluster 
phase, but they may not live long. Massive ($\ga25\,\ms$) stars formed in massive star clusters will 
probably end up as low-mass black holes, after the supernova phase. Such massive clusters will
probably survive dynamically long after the infant-mortality phase (e.g. \citealt{GoBa06}). However,
CYCs like New\,Cluster\,1 may have a total mass of only $\la100\,\ms$ (Sect.~\ref{MF}). They have  
late O and B star members (Table~\ref{tab1}), with mass in the range $10-25\,\ms$. In such cases, 
supernova explosions will probably produce $\approx1.4\,\ms$ neutron stars as residuals. Two 
supernovas in such low-mass clusters imply a sudden decrease of the total cluster mass of about 70\%, 
making CYCs instantly unstable. In relative terms, CYCs will undergo more mass loss than massive 
cluster leading to a more efficient disruption (e.g. \citealt{OldOCs}). 

Compact objects such as New\,Cluster\,1 may constitute a new class of star clusters in a young
environment, with no evident gas or dust emission. The lack of emission in CYCs is the feature
that distinguishes them from the young embedded clusters (e.g. \citealt{Hodapp94}).

\section{Concluding remarks}
\label{Conclu}

Because of its sparse stellar distribution, previous studies have considered Bochum\,1 as an OB
stellar group, an open cluster, or an association. In the present paper we investigate its nature
with field-star decontaminated 2MASS photometry and stellar radial density profiles. With an age
of $\approx3$\,Myr, the Vul\,OB1 association, which contains the young OC NGC\,6823 and the very 
compact embedded cluster Cr\,404, are used for comparisons with the young stellar content and
stellar radial density profiles present in the Bochum\,1 area.

Likewise NGC\,6823, conspicuous sequences of MS and PMS stars are present in the decontaminated 
CMD of Bochum\,1, consistent with the $\approx9$\,Myr of age. However, the RDP of Bochum\,1 is 
irregular and does not follow a King-like profile, which suggests important erosion or dispersion 
of stars from a possible primordial cluster. The MS and PMS sequences of NGC\,6823 produce 
King-like RDPs, but with an important excess of MS stars near the centre. Unlike Vul\,Ob1, which 
includes the H\,II region Sh2-86, no evident gas emission appears in Bochum\,1.

We found two new small compact young clusters in the area of Bochum\,1 (one of which may be its 
remnant core), as well as one embedded cluster. One of the small compact clusters is located $\approx8\arcmin$
to the NW of Bochum\,1 (probably the remnant core of a large previous cluster). Together with a clump of 
faint stars somewhat to the north, both have been taken as the core of an extended cluster (FSR\,911).

Available evidence shows that structurally, Bochum\,1 is not a star cluster. A possible scenario
points to a fossil remain of a star cluster, as suggested by the core (New\,Cluster\,2) and the
radially decaying stellar density profile. In this context, Bochum\,1 can be a missing link connecting 
young star cluster dissolution with the formation of low-mass OB associations. However, the processes 
that generate large OB associations such as Vul\,OB1 are yet to be explored in more detail. Probably
the difference between objects like Bochum\,1, and the star cluster NGC\,6823 arises from the mass
stored in the objects. On the other hand, the erosion of a primordial cluster is not the
only explanation for the irregular radial stellar density profile of Bochum\,1. Indeed, star formation
may occur along the border of a radially expanding density wave or ionisation front (e.g. \citealt{Soria05};
\citealt{Elme77}). In some cases, the neutral interstellar medium can be compressed by the expanding bubble
above the stability criterion against gravitational collapse, which could trigger localised star formation,
giving rise to a clumpy radial density profile.

The two compact New\,Clusters\,1 and 2, and the embedded New\,Cluster\,3, are new findings in the area 
of Bochum\,1 and FSR\,911. We show that Bochum\,1 and FSR\,911 are different objects. The optically
identified small object New\,Cluster\,1 and the large infrared detected (Fig.~\ref{fig1} and Tab.~\ref{tab6})
may be the same object. However, probably because of the lack of decontamination, \citet{FSRcat} 
overestimated dimensions of FSR\,911. The whole ensemble may turn out to be a Rosetta 
Stone to decode early-dynamical evolution processes involving embedded clusters, young compact clusters,
OCs and associations. However, to unravel all evolutionary connections we probably must wait
for Gaia's\footnote{http://www.rssd.esa.int/index.php?project=Gaia} deep proper motions.

\section*{Acknowledgements}
We thank the anonymous referee for interesting suggestions.
This publication makes use of data products from the Two Micron All Sky Survey, which is a 
joint project of the University of Massachusetts and the Infrared Processing and Analysis
Center/California Institute of Technology, funded by the National Aeronautics and Space 
Administration and the National Science Foundation. This research has made use of the WEBDA
database, operated at the Institute for Astronomy of the University of Vienna. We acknowledge 
partial support from CNPq (Brazil). 



\begin{thebibliography}{}

\bibitem[\protect\citeauthoryear{Albacete Colombo et al.}{2002}]{Albacete02}
   Albacete Colombo, J.F., Flaccomio, E., Micela, G., Sciortino, S. \& Damiani, F. 
   2007, A\&A, 464, 211

\bibitem[\protect\citeauthoryear{Barkhatova}{1957}]{Barkhatova57}
   Barkhatova, K.A. 1957, SvA, 1, 822

\bibitem[\protect\citeauthoryear{van den Bergh, Morbey \& Pazder}{1991}]{vdBMP91}
   van den Bergh, S., Morbey, C. \& Pazder, J. 1991, ApJ, 375, 594

\bibitem[\protect\citeauthoryear{Bergond, Leon \& Guilbert}{2001}]{Bergond2001}
   Bergond, G., Leon, S. \& Guilbert, J. 2001, A\&A, 377, 462

\bibitem[\protect\citeauthoryear{Bica et al.}{2006}]{GCProp}
   Bica, E., Bonatto, C., Barbuy, B. \& Ortolani, S. 2006, A\&A, 450, 105

\bibitem[\protect\citeauthoryear{Bica, Bonatto \& Camargo}{2008}]{ProbFSR}
   Bica, E., Bonatto, C. \& Camargo, D.. 2008, MNRAS, 385, 349

\bibitem[\protect\citeauthoryear{Blaauw}{1964}]{Blaauw64}
   Blaauw, A. 1964, ARA\&A, 2 , 213
   
\bibitem[\protect\citeauthoryear{Bonatto, Bica \& Girardi}{2004}]{TheoretIsoc}
   Bonatto, C., Bica, E. \& Girardi, L. 2004, A\&A, 415, 571

\bibitem[\protect\citeauthoryear{Bonatto \& Bica}{2005}]{DetAnalOCs}
   Bonatto, C. \&  Bica, E. 2005, A\&A, 437, 483

\bibitem[\protect\citeauthoryear{Bonatto \& Bica}{2006}]{BB06}
   Bonatto, C., Bica, E. 2006, A\&A, 460, 83

\bibitem[\protect\citeauthoryear{Bonatto, Santos Jr. \& Bica}{2006}]{N6611}
   Bonatto, C., Santos Jr., J.F.C. \& Bica, E. 2006, A\&A, 445, 567
   
\bibitem[\protect\citeauthoryear{Bonatto et al.}{2006}]{N4755}
   Bonatto, C., Bica, E., Ortolani, S. \& Barbuy, B. 2006, A\&A, 453, 121

\bibitem[\protect\citeauthoryear{Bonatto \& Bica}{2007a}]{OldOCs}
   Bonatto, C. \& Bica, E. 2007a, A\&A, 473, 445

\bibitem[\protect\citeauthoryear{Bonatto \& Bica}{2007b}]{BB07}
   Bonatto, C. \& Bica, E. 2007b, MNRAS, 377, 1301
   
\bibitem[\protect\citeauthoryear{Bonatto \& Bica}{2008a}]{StrucSCs}
   Bonatto, C. \& Bica, E. 2008a, A\&A, 477, 829

\bibitem[\protect\citeauthoryear{Bonatto \& Bica}{2008b}]{BB08}
   Bonatto, C. \& Bica, E. 2008b, A\&A, 485, 81

\bibitem[\protect\citeauthoryear{Bonatto, Bica \& Santos Jr.}{2008}]{PNOCs}
   Bonatto, C., Bica, E. \& Santos Jr., J.F.C. 2008, MNRAS, 386, 324

\bibitem[\protect\citeauthoryear{Brand \& Blitz}{1993}]{BrBl93}
   Brand, J. \& Blitz, L. 1993, A\&A, 275, 67
   
\bibitem[\protect\citeauthoryear{Cardelli, Clayton \& Mathis}{1989}]{Cardelli89}
   Cardelli, J.A., Clayton, G.C. \& Mathis, J.S. 1989, ApJ, 345, 245
   
\bibitem[\protect\citeauthoryear{Carraro et al.}{2006}]{Carraro06}
   Carraro, G., Janes, K.A., Costa, E. \& M\'endez, R.A. 2006, MNRAS, 368, 1078

\bibitem[\protect\citeauthoryear{Collinder}{1931}]{Collinder31}
   Collinder, P. 1931, AnLun, 2, 1

\bibitem[\protect\citeauthoryear{Dias et al.}{2002}]{Dias02}
   Dias, W.S., Alessi, B.S., Moitinho, A. \& L\'epine, J.R.D. 2002, A\&A, 389, 871

\bibitem[\protect\citeauthoryear{Dutra, Santiago \& Bica}{2002}]{DSB2002}
   Dutra, C.M., Santiago, B.X. \& Bica, E. 2002, A\&A, 383, 219
   
\bibitem[\protect\citeauthoryear{Efremov \& Elmegreen}{1998}]{EfEl98}
   Efremov, Y.N. \& Elmegreen, B.G. 1998, MNRAS, 299, 588

\bibitem[\protect\citeauthoryear{Elmegreen \& Lada}{1977}]{Elme77}
   Elmegreen, B.G. \& Lada, C.J. 1977, ApJ, 214, 725

\bibitem[\protect\citeauthoryear{Froebrich, Scholz \& Raftery}{2007}]{FSRcat}
   Froebrich, D., Scholz, A. \& Raftery, C.L. 2007, MNRAS, 374, 399
   
\bibitem[\protect\citeauthoryear{de la Fuente Marcos \& de la Fuente Marcos}{2008}]{laF08}
   de la Fuente Marcos, R. \& de la Fuente Marcos, C. 2008, AJ, 672, 342

\bibitem[\protect\citeauthoryear{Girardi et al.}{2002}]{Girardi02}
   Girardi, L., Bertelli, G., Bressan, A., Chiosi, C., Groenewegen, M.A.T.,
   Marigo, P., Salasnich, B. \& Weiss, A. 2002, A\&A, 391, 195

\bibitem[\protect\citeauthoryear{Goodwin \& Bastian}{2006}]{GoBa06}
   Goodwin, S.P. \& Bastian, N. 2006, MNRAS, 373, 752
   
\bibitem[\protect\citeauthoryear{Hodapp}{1994}]{Hodapp94}
   Hodapp, K.-W. 1994, ApJS, 94, 615

\bibitem[\protect\citeauthoryear{Kharchenko et al.}{2005}]{Kharchenko05}
   Kharchenko, N.V., Piskunov, A.E., R\"oser, S., Schilbach, E. \& Scholz, R.-D. 2005,
   A\&A, 438, 1163

\bibitem[\protect\citeauthoryear{King}{1962}]{King1962}
   King, I. 1962, AJ, 67, 471
   
\bibitem[\protect\citeauthoryear{Kn\"odlseder}{2000}]{Kn00}
   Kn\"odlseder, J.  2000, A\&A, 360, 539 

\bibitem[\protect\citeauthoryear{Lada \& Lada}{2003}]{LL2003}
   Lada, C.J. \& Lada, E.A. 2003, ARA\&A, 41, 57

\bibitem[\protect\citeauthoryear{Lamers et al.}{2005}]{Lamers05}
   Lamers, H.J.G.L.M., Gieles, M., Bastian, N., Baumgardt, H.,
   Kharchenko, N.V. \& Portegies Zwart, S. 2005, A\&A, 441, 117

\bibitem[\protect\citeauthoryear{Lyng\aa}{1982}]{Lynga82}
   Lyng\aa, G. 1982, A\&A, 109, 213

\bibitem[\protect\citeauthoryear{Maciejewski \& Niedzielski}{2007}]{MacNie07}
   Maciejewski, G. \& Niedzielski, A. 2007, A\&A, 467, 1065
   
\bibitem[\protect\citeauthoryear{Ma\'\i z-Apell\'aniz}{2001}]{MA01}
   Ma\'\i z-Apell\'aniz, J. 2001, ApJ, 560, L83

\bibitem[\protect\citeauthoryear{Massey, Johnson \& Gioia-Eastwood}{1995}]{Massey95}
   Massey, P., Johnson, K.E. \& De Gioia-Eastwood, K. 1995, ApJ, 454, 151
   
\bibitem[\protect\citeauthoryear{Mercer et al.}{20005}]{Mercer05}
   Mercer, E.P., Clemens, D.P., Meade, M.R., Babler, B.L., Indebetouw, R.,
   Whitney, B.A., Watson, C., Wolfire, M.G. et al. 2005, ApPJ, 6335,  560

\bibitem[\protect\citeauthoryear{Mermilliod \& Paunzen}{2003}]{Merm03}
   Mermilliod, J.C. \& Paunzen, E. 2003, A\&A, 410, 511

\bibitem[\protect\citeauthoryear{Moffat \& Vogt}{1975}]{Moffat75}
   Moffat A.F.J. \& Vogt N. 1975, A\&AS, 20, 85

\bibitem[\protect\citeauthoryear{Nilakshi, Pandey \& Mohan}{2002}]{Nilakshi02}
   Nilakshi, S.R., Pandey, A.K. \& Mohan, V. 2002, A\&A, 383, 153
   
\bibitem[\protect\citeauthoryear{Pigulski, Kolaczkowski \& Kopacki}{2000}]{Pigu00}
   Pigulski, A., Kolaczkowski, Z. \& Kopacki, G. 2000, AcA, 50, 113

\bibitem[\protect\citeauthoryear{Piskunov et al.}{2006}]{Piskunov06}
   Piskunov, A.E., Kharchenko, N.V., R\"oser, S., Schilbach, E. \& Scholz,
   R.-D., 2006, A\&A, 445, 545
   
\bibitem[\protect\citeauthoryear{Piskunov et al.}{2008}]{Piskunov08}
   Piskunov, A.E., Schilbach, E., Kharchenko, N.V., R\"oser, S. \& Scholz,
   R.-D., 2008, A\&A, 477, 165

\bibitem[\protect\citeauthoryear{Salpeter}{1955}]{Salpeter55}
   Salpeter, E. E. 1955, ApJ, 121, 161

\bibitem[\protect\citeauthoryear{Sharma et al.}{2006}]{Sharma06}
   Sharma, S., Pandey, A. K., Ogura, K., Mito, H., Tarusawa, K. \& Sagar, R.
   2006, AJ, 132, 1669

\bibitem[\protect\citeauthoryear{Sharpless}{1959}]{Sharpless59}
   Sharpless, S. 1959, ApJS, 4, 257

\bibitem[\protect\citeauthoryear{Siess, Dufour \& Forestini}{2000}]{Siess2000}
   Siess, L., Dufour, E. \& Forestini, M. 2000, A\&A, 358, 593

\bibitem[\protect\citeauthoryear{Soares et al.}{2005}]{Soares05}
   Soares, J.B., Bica, E., Ahumada, A.V. \& Clari\'a, J.J. 2005, A\&A, 430, 987

\bibitem[\protect\citeauthoryear{Soares et al.}{2008}]{Soares08}
   Soares, J.B., Bica, E., Ahumada, A.V., Clari\'a, J.J. 2008, A\&A, 478, 419

\bibitem[\protect\citeauthoryear{Soria et al.}{2005}]{Soria05}
   Soria, R., Cropper, M., Pakull, M., Mushotzky, R. \& Wu, K. 2005, MNRAS, 356, 12

\bibitem[\protect\citeauthoryear{Stephenson \& Sanduleak}{1971}]{StepSand71}
   Stephenson, C.B. \& Sanduleak, N. 1971, Publ. Warner \& Swasey Obs. 1, 1

\bibitem[\protect\citeauthoryear{Tadross et al.}{2002}]{Tad2002}
   Tadross, A.L., Werner, P., Osman, A. \& Marie, M. 2002, NewAst, 7, 553

\bibitem[\protect\citeauthoryear{Torres et al.}{2000}]{Torres00}
   Torres, C.A.O., da Silva, L., Quast, G.R., de la Reza, R. \& Jilinski, E.
   2000, AJ, 120, 1410

\bibitem[\protect\citeauthoryear{Trager at al.}{1995}]{Trager95}
   Trager, S.C., King, I.R., Djorgovski, S. 1995, AJ, 109, 218

\bibitem[\protect\citeauthoryear{Yadav \& Sagar}{2003}]{Yadav03}
   Yadav, R.K.S. \& Sagar, R. 2003, BASI, 31, 87

\bibitem[\protect\citeauthoryear{Yonekura et al}{2005}]{Yonekura05}
    Yonekura, Y., Asayama, S., Kimura, K., Ogawa, H., Kanai, Y., Yamaguchi,
    N., Barnes, P. J. \& Fukui, Y. 2005, ApJ, 634, 476

\label{lastpage}

\end{thebibliography}
\end{document}